\documentclass[%
 reprint,
 amsmath,amssymb,
 aps,
]{revtex4-2}

\usepackage{graphicx}
\usepackage{dcolumn}
\usepackage{bm}
\usepackage{braket}
\usepackage{mathtools}
\usepackage{tensor}
\allowdisplaybreaks

\usepackage[resetlabels,labeled]{multibib}
\newcites{App}{Appendix References}

\newcommand{\ld} {m} 
\newcommand{\nn}{\nonumber}

\begin{document}

\preprint{APS/123-QED}

\title{Statistical properties of large data sets with linear latent features}

\author{Philipp Fleig}
\affiliation{Department of Physics \& Astronomy, University of Pennsylvania, Philadelphia, PA 19104, USA}
\author{Ilya Nemenman}%
\affiliation{Department of Physics, Emory University, Atlanta, GA 30322, USA}
\affiliation{Department of Biology, Emory University, Atlanta, GA 30322, USA}
\affiliation{Initiative in Theory and Modeling of Living Systems, Atlanta, GA 30322, USA}%

\date{\today}

\begin{abstract}
 Analytical understanding of how low-dimensional latent features reveal themselves in large-dimensional data is still lacking. We study this by defining a linear latent feature model with additive noise constructed from probabilistic matrices, and analytically and numerically computing the statistical distributions of pairwise correlations and eigenvalues of the correlation matrix. This allows us to resolve the latent feature structure across a wide range of data regimes set by the number of recorded variables, observations, latent features and the signal-to-noise ratio. We find a characteristic imprint of latent features in the distribution of correlations and eigenvalues and provide an analytic estimate for the boundary between signal and noise even in the absence of a clear spectral gap.
\end{abstract}

\maketitle


{\em Introduction.}
Massively parallel experiments are now standard in all fields of science.  They record the state of the system through a large number, $N$, of variables $x_i,\, i=1\dots N$. These variables could be positions of particles, agents, or tracers  in dusty plasmas~\cite{Killer_etal_2016},  soft matter~\cite{valentine2001investigating}, insect swarming~\cite{sinhuber2019three}, and dynamical systems~\cite{Lusch_etal_2018}. Or they can be field values at different spatial points in fluids~\cite{Schanz2016ShakeTheBoxLP}, climate data~\cite{climate_data}, or activity of ``nodes''  in gene expression networks \cite{Natale:2017ej}, neural  recordings  \cite{Meshulam:2017kb}, postures \cite{stephens2008dimensionality,Berman_etal_2014}, biodiversity~\cite{WEISSER20171}, ecology~\cite{Dell_etal_2014}, etc. Crucially, the number of recorded variables is often  much larger than the true number of (latent) degrees of freedom in the system \cite{Pandarinath_etal_2018,Schwab:2014io,Morrell:2021hk}. This allows to use correlations among the measured variables to detect the latent state variables, which can then be included in models of the system \cite{gallego2017neural,Pandarinath_etal_2018,Page_etal_2020,nieh2021geometry}. Such approaches generally go under the name of {\em dimensionality reduction}. While many dimensionality reduction methods have been developed, the classic Principal Components Analysis (PCA) still reigns supreme in its widespread use. PCA assumes the recorded data to be Gaussian, and the latent features to be linear combinations of the recorded variable activities. It then calculates the variable-variable correlation matrix and  identifies the eigendirections (principal components) corresponding to the large eigenvalues of the correlation matrix as the latent features. If there is a gap between $\ld\ll N$ largest eigenvalues and the rest, then $\ld$ principal components capture a large part of the variance of all variables, achieving the dimensionality reduction.

While the number of variables may be (much) larger than the number of  latent components,  modern experiments are usually undersampled since nonstationarity precludes making measurements for a long time. Thus  the number of independent measurements, $T$, is often of the same order as $N$. Because of this, statistical fluctuations in the correlation matrix are large (in fact, the matrix is degenerate for $T\le N$), and the principal components calculated from samples often only vaguely resemble their true values, with many correlations emerging as statistical artifacts. To identify if findings of the PCA analysis can be trusted, one  refers to Random Matrix Theory~\cite{Potters_Bouchaud_2020}, and in particular the Mar\v cenko-Pastur (MP) eigenvalue density of a pure noise correlation matrix~\cite{Marchenko_Pastur1967}. Specifically, one  calculates the upper and the lower bounds of eigenvalues expected by pure chance from $T$ measurements of $N$ independent variables, and eigenvalues outside this interval (and their corresponding eigenvectors) are deemed to be statistically significant. 

This straightforward approach assumes that signal-induced correlations among the variables do not influence the spectrum of the noise-induced correlations. This has never been proven and, as we will show, is, in fact, incorrect. More generally, we are not aware of results extending the MP analysis to produce correlation eigenvalue densities when the correlations come from the sampling noise {\em and} from true, low-dimensional, latent signals, with the latter coming from some known distribution in its own turn (though see work on spiked covariance matrix models~\cite{sengupta1999distributions,Loubaton_2011,capitaine2016spectrum}). Even statistics of the entries of the correlation matrix  (rather than of its eigenvalues) have not been reported in this case. In this Letter, we fill in these gaps and calculate -- analytically and numerically -- various statistical properties of correlation matrices emerging from  data sets with low-dimensional latent feature structure. We show that the distribution of pairwise correlations and the spectra of eigenvalues of these correlation matrices carry signatures of the number of latent features, allowing one not only to choose, rigorously, which of the principal components are above the noise floor, but also to see if the overall model of latent features plus noise is a good description for a particular data set.

We do our analysis in two limits. First is the~\textit{classical statistics} limit, where the number of variables is constant (though possibly large) and the number of latent features is small,  while the number of observations grows to infinity, such that $N/T\rightarrow0$. Second is the \textit{intensive} limit, where both the number of the measured variables and the number of observations grows, such that $N/T=\text{const}$, while the number of latent features stays finite $m={\rm const}$. Both of these limits routinely happen in modern datasets. We leave considerations of the~\textit{extensive} limit, where the number of latent features grows with the number of the measured variables, to a future publication. We believe that these results are an important step in the development of analytical tools for understanding large datasets.

{\em The model and its limits.} We consider observations produced by a  random matrix model that combines latent signals and uncorrelated noise:
\begin{align}\label{eqn:X_structure}
    \mathbf X = \mathbf U \mathbf V+\sigma \mathbf R\,.
\end{align}
The component matrices of the latent signal term $\mathbf{U}$ and $\mathbf{V}$ have dimensions $T\times m$ and $m\times N$, respectively. Thus $m$ latent features get randomly sampled $T$ times (matrix $\mathbf U$), and each of the $N$ measured variables is a random linear combination of the latent features (matrix $\mathbf V$). We assume $m\leq T,N$ throughout this work, such that the  rank of the signal matrix $\mathbf U\mathbf V$ is equal to the number of latent features $\ld$. In other words, an estimate of the features can be inferred from the samples uniquely.

The entries of $\mathbf{U}$ and $\mathbf V$ are Gaussian random variables with  zero mean and variances $\sigma_U^2$ and $\sigma_{V}^2$, respectively:
\begin{align}
   & U_{t\mu} \sim \mathcal N(0,\sigma_U^2)\,,\quad V_{\mu n} \sim \mathcal N(0,\sigma_V^2)\,,\label{eq:vars}\\
    &t=1,\ldots,T,\; \mu=1,\ldots,\ld,\; n=1,\dots,N. 
\end{align}
We make this choice for analytic tractability; in applications to real data, the variances of the entries may need to be matched to variances of each measured variable. Finally, the elements of the noise matrix $\mathbf R$ are  i.i.d.\ unit variance Gaussian random variables, so that  the noise in every observation  has  variance $\sigma^2$.

From Eqs.~(\ref{eqn:X_structure}, \ref{eq:vars}), the elements of the signal matrix $\mathbf{UV}$ are a sum of $\ld$ products of two Gaussian variables with variances $\sigma_U^2$ and $\sigma_V^2$. In {\em Online Supplementary Materials~\ref{app:data_distribution}} we use characteristic functions to derive the probability density of these entries and show that their variance is 
\begin{align}
\sigma_{UV}^2=\ld\sigma_U^2\sigma_V^2\,.
\label{eq:var}
\end{align}
In other words, as expected from addition of independent random variables, each of the latent components adds $\sigma^2_U\sigma^2_V$ to the variance of the observations. 

Further, for a large number of latent features $\ld$, the probability density of ${\mathbf X}$ approaches a Gaussian with zero mean, see Fig.~\ref{fig:data_to_Gaussian}. This allows us to define a Gaussian  signal-to-noise ratio, $\textsf{SNR}\equiv\sigma_{UV}^2/\sigma^2= \ld\sigma_U^2\sigma_V^2/\sigma^2$. Since our goal is to calculate  properties of the data matrix independent of the units chosen to measure each of the variables,  we normalize  the data matrix
\begin{equation}\label{eq:X_standardised}
    \widetilde{\mathbf X}\equiv\mathbf X/\sigma_X,\; \sigma_X^2\equiv\sigma_{U V}^2+\sigma^2= \sigma^2(1+\textsf{SNR}).
\end{equation}
Note that this normalization is by an expected standard deviation, and is different from subtracting empirical means and standardizing by an empirical standard deviation. However, we expect the difference to be of order $T^{-1/2}$, and thus negligible in what follows.

In this work, we will be exploring the properties of the normalized empirical covariance matrix (NECM)
\begin{multline}
\label{eq:NECM}
{\mathbf C}=\frac{1}{T}\widetilde{\mathbf X}^T\widetilde{\mathbf X}=\frac{1}{T}(\widetilde{\mathbf U\mathbf V})^T(\widetilde{\mathbf U\mathbf V})+\tilde\sigma^2\mathbf R^T\mathbf R\\\quad+\tilde\sigma(\widetilde{\mathbf U\mathbf V})^T\mathbf R+\tilde\sigma\mathbf R^T\widetilde{\mathbf U\mathbf V}\,,
\end{multline}
as well as the matrix of correlation coefficients 
\begin{equation}
\label{eq:corr_matrix}
c_{pq}=\frac{C_{pq}}{\sqrt{C_{pp}}\sqrt{C_{qq}}}\,.
\end{equation}
To explore  different regimes of the problem, we use the following parameters: 
\begin{align}
q \equiv N/T,\; q_U\equiv\ld/T,\; \mbox{and}\; q_{V^T}\equiv\ld/N,
\end{align}
only two of which are independent. These parameters  emerge naturally in our theoretical analysis in {\em Online Supplementary Materials~\ref{app:parameterisation}} as controlling behavior of various observables in the model. Then  the classical statistics and the intensive limits, introduced above, become:
\begin{align}
       \textit{Classical stats.:}&\; &q\rightarrow0,\;q_U\rightarrow0,\;q_{V^T}=\text{const}\,,\label{eq:classic_limit}\\
    \textit{Intensive:}&\; &q=\text{const},\;q_U\rightarrow0,\;q_{V^T}\rightarrow0\label{eq:intensive}\,,
\end{align}
together with $\textsf{SNR}=\text{const}$ in both limits. Notably, $q_U^{-1}$, gives the number of observations available per latent feature to be learned. Since the parameter $q_U$ is small in both limits,  it means that the latent features are well sampled in our model, even if the measured variables (controlled by $q$) may not be.

\emph{Density of pairwise correlations.} 
\begin{figure*}
\includegraphics[width=1.\linewidth]{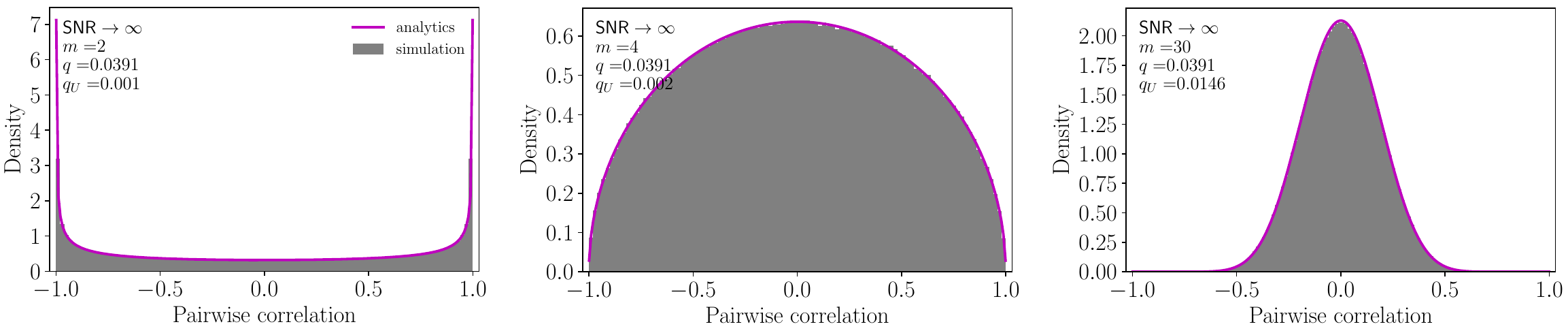}
\caption{Distribution of pairwise correlations in the pure signal limit $\textsf{SNR}\rightarrow\infty$, for $\ld=2,4$ and $30$ latent features. Analytic form given by a symmetric Beta distribution (magenta) and simulated data (gray). Each simulations is run with $N=80$ variables and $T=2048$ observations and constitutes $1000$ independent model realisations.}\label{fig:fig1}
\end{figure*}
The first observable statistics we calculate is the probability density of correlations $c_{pq}$ of the standardized variables $\widetilde{ \mathbf X}$ in our latent features model. Our goal is to analyze the dependence of the density of the matrix entries $c_{pq}$ on $\ld$, $T$, and the noise strength. The numerator in the correlation matrix in Eq.~(\ref{eq:corr_matrix}) has three contributions: $(\mathbf{UV})^T(\mathbf{UV})$ from the pure latent features signal, $\mathbf R^T\mathbf R$ from the pure noise, and two cross terms between the signal and the noise, e.~g.~$(\mathbf{UV})^T\mathbf R$. Each of these terms is analyzed separately in  {\em Online Supplementary Materials~\ref{app:pdf_corr_coeffs}}, and we reduce each term to correlations between independent Gaussian vectors. Such correlations are distributed according to the symmetric Beta distribution~\cite{Hotelling_1953}
\begin{align}\label{eq:Beta_main}
    \mathrm{pdf}(r)=\text{Beta}(r;\alpha,\alpha;\ell=-1;s=2)\,,
\end{align}
where the location $\ell$ and scale $s$ of the Beta distribution are set such that correlations fall on the interval  $[-1,1]$. The shape parameter $\alpha$ is determined individually for each contribution. For the signal-signal, noise-noise, and signal-noise contribution we find
\begin{align}
    \alpha_\text{s}=\frac{\ld-1}{2}\,,\;
    \alpha_\text{n}=\frac{T-1}{2}\,,\;
    \alpha_\text{sn}=\frac{\ld^{1/2}T^{1/2}-1}{2}\,,
\end{align}
respectively.
The sum of Beta distributions can be well approximated by a single Beta distribution~\cite{sum_iid_Beta_distributed_variables}. This finally allows us to approximate the distribution of the entries of the full correlation matrix by a single Beta distribution of the form Eq.~\eqref{eq:Beta_main}. In {\em Online Supplementary Materials~\ref{app:pdf_corr_coeffs}}, we show that in the limit when contributions of $\mathcal O(T^{-1/2})$ and $\mathcal O(\ld^{-1/2})$ can be neglected, the parameter of the approximating distribution is
\begin{align}
    \alpha\approx\frac{\left(\sqrt{\ld(1+1/\textsf{SNR})}^{-1}+\sqrt{T(1+\textsf{SNR})}^{-1}\right)^{-2}-1}{2}\,.
\end{align}
Notably, $\alpha$ is a function of $\ld$, and hence the shape of the distribution depends on $\ld$. Thus the number of latent dimensions in data can be estimated from the empirical distribution of the correlation coefficients. 

Numerical validation of this result in the pure signal limit $\textsf{SNR}\rightarrow\infty$, is shown in Fig.~\ref{fig:fig1}. For small $\ld$, the distribution distinctly changes shape as $\ld$ varies. However, when the number of latent features becomes comparable to the number of variables, and thus $q_{V^T}\rightarrow 1$, the distribution approaches a Gaussian, making it increasingly difficult to infer the precise value of $\ld$ from its shape. The quality of the analytic approximation increases the smaller $q_U$, i.~e., when more observation per latent feature are available. The analytic approximation is also exact in the large noise limit $\mathsf{SNR}\rightarrow0$. However, deviations appear for finite $\mathsf{SNR}$ when $\ld$ is small, see Fig.~\ref{fig:correlations_finitesmallSNR}.

\emph{Eigenvalue density.}
We compute the eigenvalue density of NECM, $\mathbf C$, cf.~Eq.~\eqref{eq:NECM}, from its Stieljtes transform, $g_N(z)=N^{-1}\text{Tr}(z\mathbf I-\mathbf C)^{-1}$, where $z$ is a complex number and denote the large-$N$ limit of $g^N_{\mathbf C}$ by $\mathfrak g_\mathbf{C}$~\cite{Potters_Bouchaud_2020}. The density of eigenvalues is then obtained from the Sokhotski–Plemelj formula
\begin{align}\label{eq:sokhotski-plemelj}
    \rho(\lambda)=\frac{1}{\pi}\lim_{\eta\rightarrow 0^+}\Im \mathfrak g_\mathbf{C}(z=\lambda-i\eta)\,,
\end{align}
where $\Im$ denotes the imaginary part. 

\begin{figure*}
\includegraphics[width=0.9\linewidth]{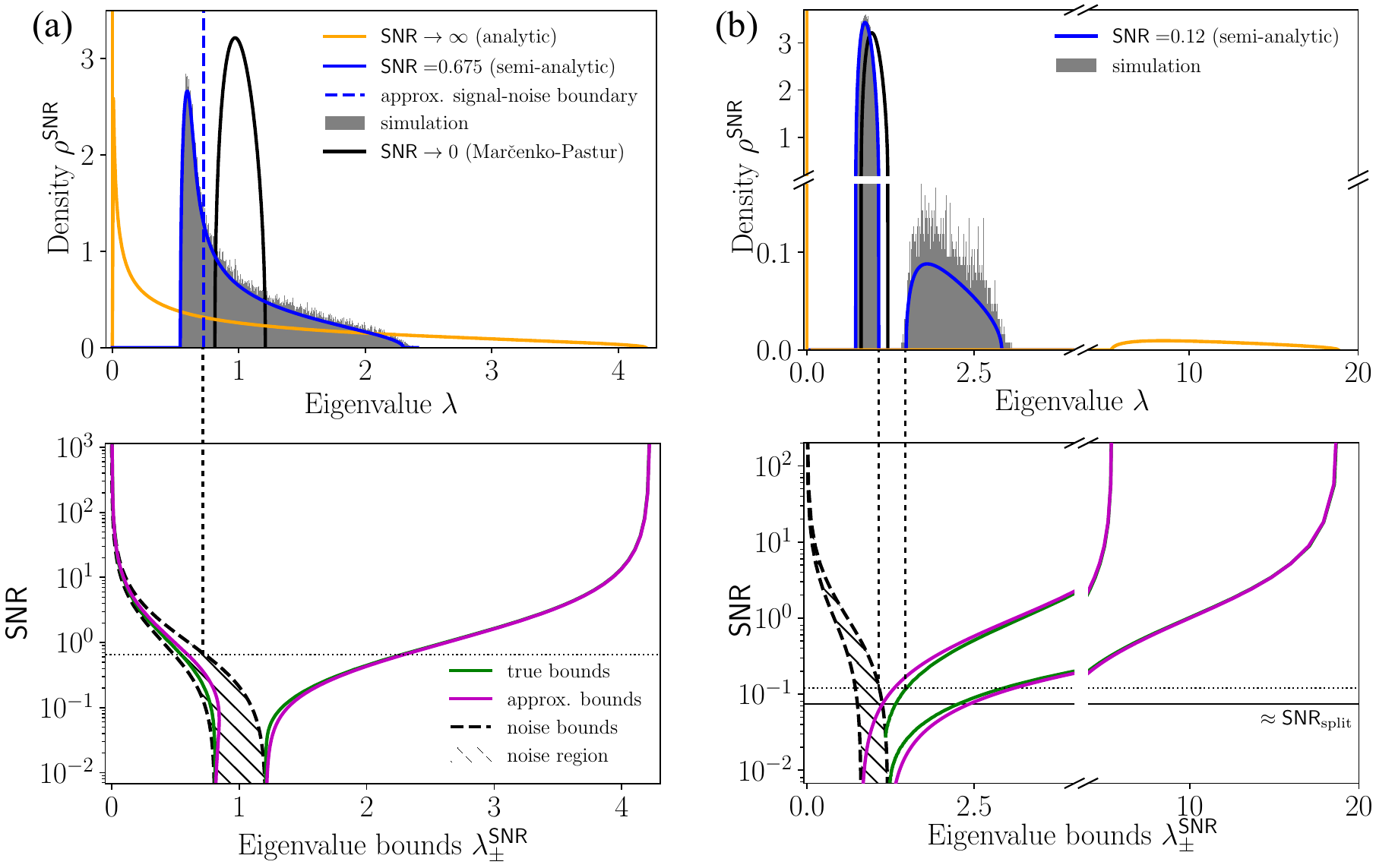}
\caption{Eigenvalue density and bounds as a function of the $\textsf{SNR}$ in the classical statistics and intensive limit. (a) \textit{Classical statistics} limit ($q=0.01$, $q_{V^T}=0.9$) at three levels of $\textsf{SNR}$. Top plot: zero noise analytic density Eq.~\eqref{eq:classical_density_SNRinf} (yellow); large noise limit given by the Mar\v cenko-Pastur distribution (black); and intermediate noise semi-analytic density (blue) with the numerical simulation (gray) for comparison. Vertical dashed blue line is the approximate boundary between noise (left) and signal (right). Bottom: eigenvalue bounds $\lambda_\pm^\textsf{SNR}$, as a function of the $\textsf{SNR}$. True bounds obtained as numerical solution of Eq.~\eqref{eq:classical_polynomial_wnoise} (green) and approximate bounds given by the analytic expression in Eq.~\eqref{eq:bound_approx} (magenta). The approximate noise region is striped. Horizontal dotted line indicates the $\mathsf{SNR}$ value of the blue density curve and the dashed line extending across plots indicates the signal-noise boundary. (b) \textit{Intensive} limit ($q=0.01$, $q_{V^T}=0.09$) with plots analogous to (a). Top: zero noise density Eq.~\eqref{eq:intensive_density_SNRinf} (yellow); for intermediate noise (blue), the left bump corresponds to noise and the right bump to latent feature signal. Bottom: for $\textsf{SNR}\gtrapprox\textsf{SNR}_\text{split}$ (horizontal solid black line), the density splits into two bumps. Simulations constitute $360$ independent realisations of the model with $N=300$ and $\sigma_U^2\sigma_V^2=1$.}\label{fig:fig2}
\end{figure*}

Full details of the computations are given in~{\em Supplementary Online Materials~\ref{app:eigenvalue_density}}. Briefly, we consider the eigenvalue density in the classical and intensive limits,  Eqs.~(\ref{eq:classic_limit},~\ref{eq:intensive}). This allows us to simplify the calculations by neglecting the cross terms between the signal $\mathbf{UV}$ and the noise $\mathbf R$, cf.~{\em Supplementary Online Materials~\ref{app:signal_noise_approx}}. Then, in the \textit{classical statistics} limit, the Stieltjes transform satisfies the third order polynomial equation
\begin{align}\label{eq:classical_polynomial_wnoise}
    a\mathfrak g_{\mathbf C}^3+b\mathfrak g_{\mathbf C}^2+c\mathfrak g_{\mathbf C}+d=0\,,
\end{align}
with coefficients
\begin{align}
    a&=\frac{qz}{1+\textsf{SNR}}\,,\\
    b&=-\frac{qq_{V^T}z}{\textsf{SNR}} + \frac{(q_{V^T} - 1)q + 1}{1+\textsf{SNR}} - z\,,\label{eq:classical_bcoeff}\\
    c&=\frac{(q - 1)q_{V^T}}{\textsf{SNR}} + q_{V^T}z\big(1+\textsf{SNR}^{-1}\big) - q_{V^T} + 1\,,\\
    d&=-q_{V^T}\big(1+\textsf{SNR}^{-1}\big)\,.
\end{align}
While one can solve this cubic equation analytically, the resulting equations are unwieldy, allowing for little direct insights. Therefore we rely on analytic approximations in the two noise limits as well as numerics. Taking the zero noise limit, $\textsf{SNR}\rightarrow\infty$, the equation reduces to a quadratic polynomial which we solve and evaluate Eq.~\eqref{eq:sokhotski-plemelj}, to find the exact eigenvalue density
\begin{align}\label{eq:classical_density_SNRinf}
    \rho^{\infty}(\lambda)&=\frac{\sqrt{(\lambda-\lambda_-^{\infty})(\lambda_+^{\infty}-\lambda)}}{2 \pi \lambda{\bar\sigma_X}^{-2}q_{V^T}^{-1}}+(1-q_{V^T})\delta(\lambda)\,,
\end{align}
where $\bar\sigma_X^2\equiv\sigma_X^2/\ld=\sigma_U^2\sigma_V^2$. The delta function represents the $N-\ld$ eigenvalues of the NECM which are trivially zero. The $\ld$ non-trivial eigenvalues lie in a finite interval with bounds
\begin{align}\label{eq:classical_bound_SNRinf}
    \lambda_{\pm}^{\infty}=
        \bar\sigma_X^{-2}\left(1\pm \sqrt{q_{V^T}}^{-1}\right)^2\,.
\end{align}
The eigenvalue density vanishes everywhere else. For finite $\textsf{SNR}$ we solve Eq.~(\ref{eq:classical_polynomial_wnoise}) numerically.
A comparison of eigenvalue densities in the different noise regimes, including the MP density~\cite{Marchenko_Pastur1967} representing the pure noise limit for $\textsf{SNR}\rightarrow0$, is shown in the top plot of Fig.~\ref{fig:fig2}(a).

In {\em Online Supplementary materials~\ref{app:classical_limit_noise}}, we derive an analytic approximation for the eigenvalue bounds, $\lambda^\textsf{SNR}_\pm$, at finite $\textsf{SNR}$, given by a weighted average of the zero noise bounds, $\lambda_\pm^{\infty}$, and those of the MP density $\lambda_\pm^\text{MP}=(1\pm\sqrt{q})^2$ 
\footnote{Mar\v cenko-Pastur (MP) distribution, $\rho^\text{MP}(\lambda)=\sqrt{(\lambda-\lambda^\text{MP}_-)(\lambda^\text{MP}_+ -\lambda)}/(2\pi q\lambda)$, with $\lambda^\text{MP}_\pm=(1\pm\sqrt{q})^2$,~\cite{Potters_Bouchaud_2020}.}:
\begin{align}\label{eq:bound_approx}
    \lambda^\textsf{SNR}_\pm&\approx\big(1+\textsf{SNR}^{-1}\big)^{-1}\, \lambda_\pm^{\infty}+\big(1+\textsf{SNR}\big)^{-1}\,\lambda_\pm^\text{MP}\,.
\end{align}
In the bottom plot of Fig.~\ref{fig:fig2}(a), we show a comparison between this approximation and the true, numerically computed, bounds at different values of the $\textsf{SNR}$. The approximation is good everywhere, with the largest deviation at $\textsf{SNR}\sim10^{-1}$. The magnitude of this value is a result of the particular choice we have made to define the $\textsf{SNR}$. The part of the eigenvalue spectrum associated with the pure latent feature signal lies outside of the interval, $(1+\textsf{SNR})^{-1}\times[\lambda_-^\text{MP},\lambda_+^\text{MP}]$. Eigenvalues within this interval, shown as a striped band, correspond to noise. Notice that as the $\textsf{SNR}$ is increased the noise range is shifted to the left compared to the MP range due to the presence of the latent features signal renormalizing the NECM. Thus using the na\"ive MP bounds for rejection of eigenvalues as noise -- a common procedure in data analysis -- may be a bad practice. 

A different picture emerges in the~\textit{intensive} limit. Here the polynomial equation for $\mathfrak g_{\mathbf C}$ is of the sixth order with lengthy expressions for the polynomial coefficients, cf.~Eq.~\eqref{eq:intensive_polynomial}. For the $\textsf{SNR}\rightarrow\infty$ limit, we find the following analytic expression for the density
\begin{align}\label{eq:intensive_density_SNRinf}
    \rho^{\infty}(\lambda)&=\frac{\sqrt{(\lambda-\lambda_-^{\infty})(\lambda_+^{\infty}-\lambda)}}{2 \pi \lambda{\bar\sigma_X}^{-2}(1+q)q_{V^T}^{-1}}+(1-q_{V^T})\delta(\lambda)\,,
\end{align}
with the eigenvalue bounds given by:
\begin{align}\label{eq:intensive_bound_SNRinf}
    \lambda_{\pm}^{\infty}
    =\bar\sigma_X^{-2}\left( \sqrt{1+q}\pm \sqrt{q_{V^T}}^{-1}\right)^2\,.
\end{align}
The form of the eigenvalue density at different levels of noise is shown in the top plot of Fig.~\ref{fig:fig2}(b). For large values of the $\textsf{SNR}$, there is a gap in the density between the eigenvalues corresponding to noise, and those corresponding to the signal. The gap closes at lower $\textsf{SNR}$, and the combined density  converges to the MP density for $\textsf{SNR}\rightarrow0$. As for the classical limit, the approximate expression Eq.~\eqref{eq:bound_approx} for the bounds of the signal part of the density is good, showing the largest deviation for $\textsf{SNR}\sim10^{-1}$, as shown in the bottom plot of Fig.~\ref{fig:fig2}(b). In {\em Online Supplementary material~\ref{app:intensive_limit_noise}} we estimate the $\textsf{SNR}$ value at which the eigenvalue density splits to be $\textsf{SNR}_\text{split}\approx(\lambda_+^\text{MP}-\lambda_-^\text{MP})/\lambda_-^{\infty}$.
The existence of the gap suggests that having more observed variables (and hence more data to define the latent components) makes it easier to distinguish the signal from noise. This suggests that high throughput data sets, where individual variables cannot be well sampled ($T<N$), might still be very valuable, as long as $m\ll N,T$.

\emph{Discussion.}
We calculated statistical properties of data with latent linear features, including the density of pairwise correlations, and density of eigenvalues of the NECM. Our analysis provides two important insights. First, by looking at the distribution of the correlations and their eigenvalues, one can understand whether the latent features model is a reasonable model for the data at hand, and also ballpark the number of the latent components. Importantly, even if the eigenvalue density does not have a prominent gap, one can understand that the underlying model has a latent structure, which manifests as a distortion of the MP sea of eigenvalues. This is because our signal matrix in Eq.~\eqref{eqn:X_structure} is stochastic, in contrast to spiked covariance models, where deterministic perturbations appear as delta functions in the spectrum and  are detectable only as true outliers~\cite{sengupta1999distributions,Loubaton_2011,capitaine2016spectrum}. Second, since the spectrum of the noise correlations in the latent features model is shifted compared to the MP model, one should not use a simple truncation at the right edge of the MP density to distinguish which of the principal components are statistically significant.

We note that a lot of ink has been expended to decide whether a few variables measured well many times are more or less valuable than many variables measured infrequently and with high noise. We find that, in the intensive limit, the many measured variables lead to separation of the noise and the signal eigenvalues, resulting in a potentially more accurate discrimination of noise from signal. However, the number of observations, and their quality contribute to the $\textsf{SNR}$, a high value of which is also required for the opening of the signal-noise gap. Thus the quality and the quantity of measurements all contribute to the value data in very specific ways, which we now understand.

Our analysis involved a few approximations. The strongest of these was in neglecting the signal-noise contributions in the computation of the eigenvalue density. Including these contributions would make the polynomial equation for the Stieltjes transform substantially more complicated. However, we do not expect significant qualitative changes to the structure of the eigenvalue density in the limits considered. An additional limitation  is that the approximation for the bounds of the eigenvalue density, Eq.~\eqref{eq:bound_approx}, is strictly only valid in the extreme noise limits, $\textsf{SNR}\rightarrow0$ and $\textsf{SNR}\rightarrow\infty$. In deriving the analytic density of correlations, we assumed $q_U\rightarrow0$ in accordance with the limits in Eq.~\eqref{eq:classic_limit} and Eq.~$\eqref{eq:intensive}$, and worked to leading order in $T$.  We expect the quality of the analytic density to improve if these assumptions are removed, in particular in the regime of finite $\mathsf{SNR}$, cf.~Fig.~\ref{fig:correlations_finitesmallSNR}.

Finally, to connect our results with the analysis of real data, additional steps are required.  To model data, the means and variances of rows and columns of our matrix model have to be fit to the data. Similarly, methods to estimate the $\textsf{SNR}$ from the data and to determine whether the Gaussian assumption for the distribution of the noise and the latent components is valid will need to be developed.  Some of such extension may remain analytically tractable. Further, for our model and its extensions, it is also important to calculate the expected overlap of empirical eigenvectors with their true values.

\begin{acknowledgments}
 We are grateful to Marc Potters for his insightful comments. IN thanks the Aspen Center for Physics, partially funded by NSF Grant PHY-1607611, for hospitality. PF thanks Mirna Kramar for continued discussions and support. This work was supported in part by the Simons Foundation Grants 400425 (PF) and 827661 (IN) and by NSF Grants BCS-1822677, PHY-2014173, and PHY-2010524 (IN).

\end{acknowledgments}

\bibliography{apssamp}

\appendix
\renewcommand\thefigure{S\arabic{figure}}    
\setcounter{figure}{0}    

\section{Data distribution for the latent feature model with no noise, its variance and large $\ld$ limit}\label{app:data_distribution}
Each entry $X_{ij}$ of the latent features data matrix $\mathbf U\mathbf V$ is given by the sum of $\ld$ products of two i.i.d. Gaussian random variables
$u\sim\mathcal N(0,\sigma_U^2)$ and $v\sim\mathcal N(0,\sigma_V^2)$:
\begin{align}
    X_{ij}\sim \sum_{\mu=1}^\ld uv\,.
\end{align}
The product, $x=uv$, is distributed according to the normal product distribution~\cite{wishart_bartlett_1932_App}:
\begin{align}
    x\sim \frac{K_0\left(\frac{|x|}{\sigma_U\sigma_V}\right)}{\pi\sigma_U\sigma_V}\,,
\end{align}
where $K_\nu$ is the modified Bessel function of the second kind:
\begin{align}\label{eq:BesselFunction2ndKind}
    K_\nu(x)=\frac{\Gamma\left(\nu+\frac12\right)(2x)^\nu}{\sqrt{\pi}}\int_0^\infty d q\frac{\cos(q)}{(x^2+q^2)^{\nu+1/2}}\,.
\end{align}

To derive the probability density of the latent feature model entries $X_{ij}$,  we first compute the characteristic function $\varphi_x$ by taking the Fourier transform of the normal product distribution. We then use the fact that the characteristic function $\varphi_X$ of the sum of $\ld$ products $x$ is given by
$\varphi_{X}=\left(\varphi_x\right)^\ld$. The inverse Fourier transform of $\varphi_X$ then yields the sought after probability density.

Specifically, the characteristic function $\varphi_x$ of the normal product distribution is
\begin{align}
    \varphi_x(t)&=\mathbb E(e^{itx})=\int_{-\infty}^{\infty}dx\, \frac{K_0\left(\frac{|x|}{\sigma_U\sigma_V}\right)}{\pi\sigma_U\sigma_V} e^{itx}\nn\\
    &=\int_{-\infty}^{\infty}dx\, \frac{K_0\left(|x|\right)}{\pi} e^{it\sigma_U\sigma_Vx}\nn\\
    &=\frac{1}{\pi}\int_{-\infty}^{\infty}dx \,\int_0^\infty dq\,\frac{\cos(q)}{\sqrt{|x|^2+q^2}}e^{it\sigma_U\sigma_Vx}\nn\\
        &=\frac{1}{\pi}\int_{-\infty}^{\infty}dx \,\int_0^\infty dq\,\frac{\cos(xq)}{\sqrt{1+q^2}}e^{it\sigma_U\sigma_Vx}\nn\\
    &= \frac{1}{\pi}\int_{0}^{\infty}dq\frac{1}{\sqrt{1+q^2}}\nn\\
    &\quad\int_{-\infty}^\infty \frac{dx}{2\pi} \,\left(e^{ix(q+\sigma_U\sigma_Vt)}+e^{ix(\sigma_U\sigma_Vt-q)}\right)\nn\\
    &= \int_{0}^{\infty}dq\frac{1}{\sqrt{1+q^2}}\left(\delta(\sigma_U\sigma_Vt+q)+\delta(\sigma_U\sigma_Vt-q)\right)\nn\\
    &= \frac{1}{\sqrt{1+\sigma_U^2\sigma_V^2t^2}}
\end{align}
for $t\in\mathbb R\setminus\{0\}$, and $\delta(\cdot)$ is the Dirac delta function.

The characteristic function $\varphi_X$ of the sum of $\ld$ products $x$ is given by
\begin{align}
    \varphi_{X}=\left(\varphi_x\right)^\ld=\left(1+\sigma_U^2\sigma_V^2t^2\right)^{-\frac{\ld}{2}}
\end{align}
Finally, performing the inverse transformation we obtain the probability density function of the sum
\begin{align}
   {\rm  pdf}(X)&=\int_{-\infty}^{\infty}\frac{dt}{2\pi} \,e^{-itX}\frac{1}{\left(1+\sigma_U^2\sigma_V^2t^2\right)^{\frac{\ld}{2}}}\nn\\
    &=\int_{-\infty}^{\infty}\frac{dt}{2\pi} \,e^{-itX}\nn\\
    &\quad\int_0^\infty dq\,\frac{\delta(\sigma_U\sigma_V t+q)+\delta(\sigma_U\sigma_V t-q)}{\left(1+q^2\right)^{\frac{\ld}{2}}}\nn\\
    &=\frac{1}{\sigma_U\sigma_V}\int_0^\infty dq\nn\\
    &\quad\int_{-\infty}^{\infty}\frac{dt}{2\pi} \,e^{-\frac{itX}{\sigma_U\sigma_V}}\frac{\delta( t+q)+\delta( t-q)}{\left(1+q^2\right)^{\frac{\ld}{2}}}\nn\\
    &=\frac{1}{\pi\sigma_U\sigma_V}\int_0^\infty dq\frac{\cos(\frac{q|X|}{\sigma_U\sigma_V})}{\left(1+q^2\right)^{\frac{\ld}{2}}}\nn\\
    &=\left[\frac{|X|}{\sigma_U\sigma_V}\right]^{\ld-1}\frac{1}{\pi\sigma_U\sigma_V}\int_0^\infty dq\frac{\cos(q)}{\left(\frac{|X|^2}{\sigma_U^2\sigma_V^2}+q^2\right)^{\frac{\ld}{2}}}\nn\\
    &=\left[\frac{|X|}{2}\right]^{\frac{\ld-1}{2}}\frac{K_{\frac{\ld-1}{2}}\left(\frac{|X|}{\sigma_U\sigma_V}\right)}{(\sigma_U\sigma_V)^{\frac{\ld+1}{2}}\sqrt{\pi}\Gamma\left(\frac{\ld}{2}\right)}.\label{eq:pdf_X}
\end{align}
Since the probability density function of $X$ is symmetric around zero, the mean of the distribution vanishes:
\begin{align}
    \mu_{{UV}} &= \int_{-\infty}^\infty dX X {\rm pdf}(X)=0\,.
\end{align}
\begin{figure}
\includegraphics[width=0.95\linewidth]{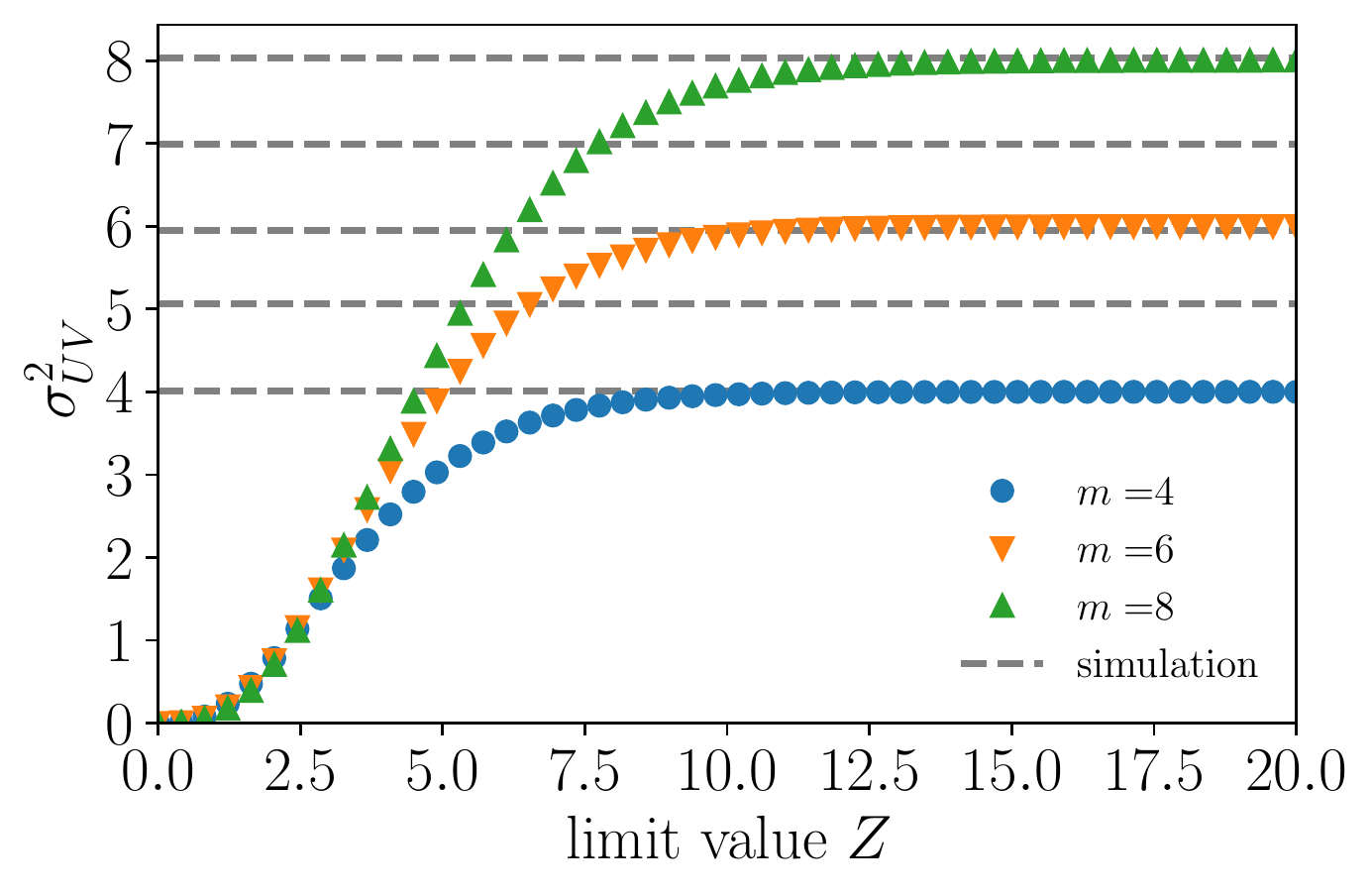}
\caption{Comparison between the variances computed from simulated data of the latent feature model for $\ld=4,5,6,7,8$ (dashed gray lines) and the numerically evaluated limit expression for $\sigma_{UV}^2$ in Eq.~\eqref{eq:variance_limit} as a function of the limit value $Z$, for even $\ld$ values. We have chosen $\sigma^2_U=\sigma^2_V=1$.}\label{fig:model_variance}
\end{figure}
The variance is 
\begin{align}
    \sigma_{{UV}}^2 &= \int_{-\infty}^\infty dX X^2 {\rm pdf}(X)\nn\\
    &=\frac{1}{2^{\frac{\ld-1}{2}}\sqrt{\pi}\Gamma\left(\frac{\ld}{2}\right)}\times\nn\\
    &\quad\times\int_{-\infty}^\infty \frac{dX}{\sigma_U\sigma_V}\left[\frac{|X|}{\sigma_U\sigma_V}\right]^{\frac{\ld-1}{2}}|X|^2K_{\frac{\ld-1}{2}}\left(\frac{|X|}{\sigma_U\sigma_V}\right)\nn\\
    &=\frac{2\sigma_U^2\sigma_V^2}{2^{\frac{\ld-1}{2}}\sqrt{\pi}\Gamma\left(\frac{\ld}{2}\right)}\int_0^\infty dX|X|^{\frac{\ld+3}{2}}K_{\frac{\ld-1}{2}}(|X|)\,.
\end{align}
The integral above can be evaluated in terms of generalized hypergeometric functions~\cite{hypergeom_integral_App}. We present the calculation for when $m$ is even in detail:
\begin{align}
    \int_0^\infty dX X^{\alpha-1}K_\nu(X)=
    \left[\Sigma(\nu,\alpha;Z)+\Sigma(-\nu,\alpha;Z)\right]_0^\infty,
\end{align}
where
\begin{align}
    \Sigma(\nu,\alpha;Z)&\equiv-\frac{2^{\nu-1}\pi Z^{\alpha-\nu}\text{csc}(\pi\nu)}{(\nu-\alpha)\Gamma(1-\nu)}\times\nn\\
    &\phantom{}_1F_2\left(\frac{\alpha-\nu}{2};1-\nu,\frac{\alpha-\nu}{2}+1;\frac{Z^2}{4}\right)
\end{align}
with parameters
\begin{align}
    \alpha\equiv\frac{\ld+5}{2}\text{  and  }\nu\equiv\frac{\ld-1}{2},
\end{align}
and $\phantom{}_1F_2$ is the generalized hypergeometric function
\begin{align}
    \phantom{}_1F_2(a_1;b_1,b_2;z)= \sum_{k=0}^\infty \frac{(a_1)z^k}{(b_1)_k(b_2)_kk!}\,.
\end{align}
In the expression above, $(\cdot)_k$ is the Pochhammer symbol, and $\text{csc}(\cdot)$ is the cosecant. Since $m$ is even, we also have $\nu\notin\mathbb Z$.
Putting everything together, we obtain the following expression for the variance
\begin{align}
    \sigma_{{UV}}^2 &=\frac{\left[\Sigma(\nu,\alpha;Z)+\Sigma(-\nu,\alpha;Z)\right]_{0}^{\infty}}{2^{\frac{\ld-3}{2}}\sqrt{\pi}\Gamma\left(\frac{\ld}{2}\right)\sigma_U^{-2}\sigma_V^{-2}}\nn\\
    &=\lim_{Z\rightarrow\infty}\frac{\Sigma(\nu,\alpha;Z)+\Sigma(-\nu,\alpha;Z)}{2^{\frac{\ld-3}{2}}\sqrt{\pi}\Gamma\left(\frac{\ld}{2}\right)\sigma_U^{-2}\sigma_V^{-2}}\,,
    \label{eq:variance_limit}
\end{align}
where we have used the fact that the numerator after the first equality vanishes at $Z=0$. 
We can evaluate the limit $Z\rightarrow \infty$, on the right-hand side numerically as shown in Fig.~\ref{fig:model_variance} and find that the variance of the latent feature data values is
\begin{align}\label{eq:UV_variance}
    \sigma_{UV}^2=m\sigma_U^2\sigma_V^2\,.
\end{align}
This is in agreement with the intuition that every latent dimension contributes its own variance to the variance of the data.

\begin{figure*}
\includegraphics[width=0.8\linewidth]{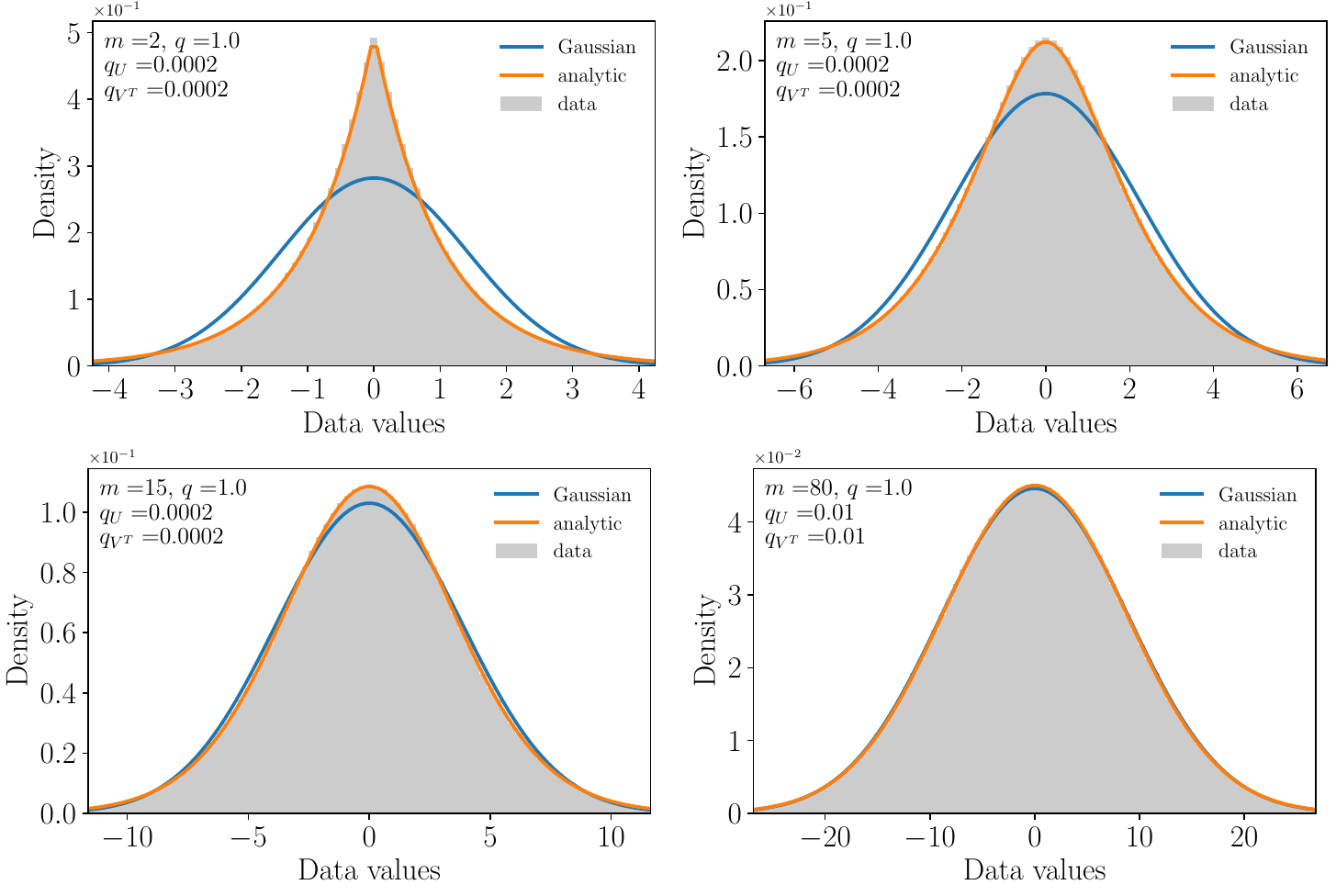}
\caption{Comparison of simulated data (gray) and the analytical distribution (orange). In the limit of large $\ld$, the distribution approaches a Gaussian form (blue). Simulated data constitutes a single realization of the model with $\sigma_U^2=\sigma_V^2=1$. \label{fig:data_to_Gaussian}}
\end{figure*}

We note that, for large values of the number of latent features $\ld$, the distribution~\eqref{eq:pdf_X} becomes normal, in agreement with the law of large numbers:
\begin{align}
    {\rm pdf}(X_{ij})=\frac{1}{\sqrt{2\pi\sigma_{UV}^2}}e^{-\frac{X^2}{2\sigma_{UV}^2}}\,.
\end{align}
Crucially, the variance of $X_{ij}$ remains $\ld$-dependent. Figure~\ref{fig:data_to_Gaussian} for compares exact analytical expression of the probability distribution and its Gaussian approximation to numerical simulations.

As a final note, if we were interested in the  distribution of data with noise, we would need to convolve the density in Eq.~(\ref{eq:pdf_X}) with the Gaussian density of the noise.

\section{Probability density of the correlation coefficients}\label{app:pdf_corr_coeffs}
For our latent features model with noise, here we calculate the probability distribution of entries in the empirical data correlation matrix. Before doing this, a few notes are in order.  
First, the correlations depend on the basis, in which variables are measured, becoming a diagonal matrix in the special case when the measured variables are the principal axes of the data cloud. Thus to make statements independent of the basis, we consider the distribution of typical correlations, or correlations in the basis random w.\ r.\ t.\ the principal axes of the data. For a given realization, the $N$-dimensional data cloud is typically anisotropic, with $\ld<N$ long directions dominated by the latent feature signal and $N-\ld$ short directions dominated by noise. When $N\gg\ld$, principal axes of the data cloud do not align with the measured variables for the vast majority of random rotations, and correlations between any random pair of variables have contributions from all latent dimensions. Thus we expect the  number of latent dimensions to be imprinted in the distribution of the elements of the correlation matrix, so that the statistics of the elements carries information about the underlying structure of the model.

\subsection{Preliminaries: Density of the correlation coefficient of two random Gaussian variables}
The correlation coefficient of two independent zero-mean variables $x$ and $y$ sampled $T$ times is
\begin{align}
    r=\frac1T\sum_t \frac{x_ty_t}{\sigma_x\sigma_y}\,,
\end{align}
where the vectors' components are mutually independent, i.i.d. random variables.  The correlation coefficient is distributed according to~\cite{Hotelling_1953}
\begin{align}
    \mathrm{pdf}(r)=\frac{\Gamma\left(\frac {T}{2}\right)}{\Gamma\left(\frac12\right)\Gamma\left(\frac{T-1}{2}\right)}(1-r^2)^{\frac{T-3}{2}}\,.
\end{align}
This can be rewritten in terms of a Beta distribution
\begin{align}
    \text{Beta}(x;\alpha,\beta)=\frac{1}{\text{B}(\alpha,\beta)}x^{\alpha-1}(1-x)^{\beta-1}\,,
\end{align}
where $x\in[0,1]$ and $\text{B}(\alpha,\beta)$ is the Beta function. Specifically,  the density of correlations is given by the symmetric Beta distribution
\begin{align}\label{eq:symmetric_Beta}
    \mathrm{pdf}(r)=\text{Beta}\left(r;\alpha,\alpha;\ell=-1,s=2\right)\,,
\end{align}
where the location $\ell$ and scale $s$ are set such that the density is defined on the interval of correlation values [-1,1], and
\begin{align}
    \alpha=\frac{T-1}{2}\,.
\end{align}
We also note that the variance of a symmetric Beta distribution with the scale $s=2$ is
\begin{align}\label{eq:var_alpha_relation}
    \text{var}=\frac{s^2}{4(2\alpha+1)}=\frac{1}{2\alpha+1}\,.
\end{align}

\subsection{Density of correlations in the latent feature model}
\begin{figure*}
\includegraphics[width=1.\linewidth]{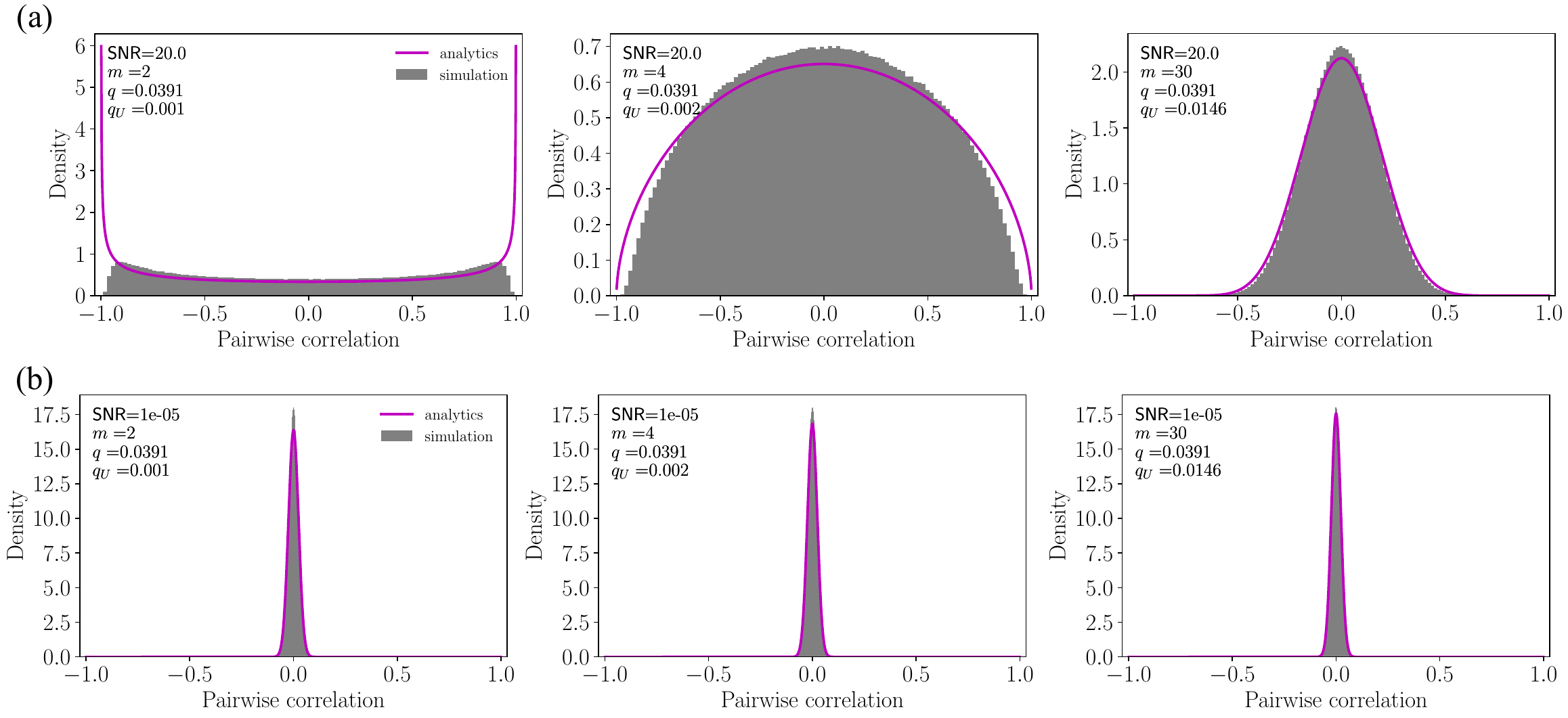}
\caption{Distribution of pairwise correlations in the regimes of finite and small signal-to-noise ratio with $\ld=2,4$ and $30$ latent features. Analytic form (magenta) and simulated data (gray). (a) $\mathsf{SNR}=20$ and (b) $\mathsf{SNR}=10^{-5}$ (large noise limit). Each simulation is run with $N=80$ variables and $T=2048$ observations and constitutes $1000$ independent model realisations.}\label{fig:correlations_finitesmallSNR}
\end{figure*}
There are multiple contributions to the correlations among the measured variables. We compute them individually, and then combine the contributions. We find that each contribution is distributed according to a symmetric Beta distribution. To obtain the overall density, we approximate the sum of Beta distributions by a single Beta distribution, the parameter of which is obtained by matching the variance to the sum of the variances of the individual components. To perform these analyses, we only keep terms to the leading order in the $\textsf{SNR}\rightarrow0$ or the $\textsf{SNR}\rightarrow\infty$ limit. Further, we assume that $q_U$ is small in accordance with the classical and intensive regimes limits.

We start with the pure noise contribution to the correlations
\begin{align}\label{eq:C_R}
    (c_{R})_{pq}&=\frac{1}{T}\sum_t\frac{ R^T_{pt}R_{tq}}{\sigma_p^\mathrm{n}\sigma_q^\mathrm{n}}\,,\\
    ({\sigma_q^\mathrm{n}})^2&=\frac{1}{T}\sum_t R^T_{qt}R_{tq}\,.
\end{align}
The expression on the right-hand side is  the correlation coefficient between two random Gaussian variables. Using  Eq.~\eqref{eq:symmetric_Beta}, we arrive at
\begin{align}\label{eq:Beta_noise}
    \mathrm{pdf}((c_{R})_{pq})=\text{Beta}\left((c_{R})_{pq};\alpha_\text{n},\alpha_\text{n};-1,2\right),\; p\neq q,
\end{align}
with
\begin{align}
    \alpha_\text{n}=\frac{T-1}{2}\,.
\end{align}
and the variance of this density is
\begin{align}
    \text{var}_\text{n}=T^{-1}\,.
\end{align}

Next we compute the density of the pure signal contribution
\begin{align}\label{eq:C_UV}
    (c_{UV})_{pq}&=\frac{1}{T\sigma^\mathrm{s}_p\sigma^\mathrm{s}_q}\sum_t \left(\sum_\mu V_{p\mu}U_{\mu t}\right)\left(\sum_\nu U_{t\nu}V_{\nu q}\right)\,,\\
    (\sigma^\mathrm{s}_p)^2 &= \frac{1}{T}\sum_t \left(\sum_\mu V_{p\mu}U_{\mu t}\right)\left(\sum_\nu U_{t\nu}V_{\nu p}\right)\,,
\end{align}
and similarly for $\sigma_q$. Rearranging, we find
\begin{align}
    (c_{UV})_{pq}&=\frac{1}{\sigma^\mathrm{s}_p\sigma^\mathrm{s}_q}\sum_{\mu\nu} V_{p\mu}V_{\nu q}\left(\frac1T\sum_t U_{\mu t}U_{t\nu}\right)\,,\\
    (\sigma^\mathrm{s}_p)^2&=\sum_{\mu\nu} V_{p\mu}V_{\nu p}\left(\frac1T\sum_t U_{\mu t}U_{t\nu}\right)\,.
\end{align}
The expression in parentheses of both of the equations above is a (co)-variance of Gaussian random numbers. For $\mu=\nu$, it follows the scaled $\chi^2$-distribution with $T$ degrees of freedom. For $\mu\neq\nu$, it is given by a rescaled version of the distribution in Eq.~(\ref{eq:pdf_X}), with $T$ instead of $m$. Crucially, the variance of either is $1/T$. Thus in the limit $q\to0$, the terms in parentheses  are $\sigma^2_U\delta_{\mu\nu}+\mathcal{O}(T^{-1/2})$, where the correction $\mathcal{O}(T^{-1/2})$ is probabilistic, but will be neglected in what follows. We get
\begin{align}
    (c_{UV})_{pq}&=\sigma_U^2\sum_{\mu\nu} \frac{V_{p\mu}V_{\nu q}\delta_{\mu\nu}}{\sigma^\mathrm{s}_p\sigma^\mathrm{s}_q}=\ld \sigma_U^2 \left(\frac{1}{\ld}\sum_{\mu} \frac{V_{p\mu}V_{\mu q}}{\sigma^\mathrm{s}_p\sigma^\mathrm{s}_q}\right) \,,\\
    (\sigma^\mathrm{s}_p)^2&=\ld\sigma_U^2 \left(\frac{1}{\ld}\sum_{\mu} V_{p\mu}^2\right)\,,
    \label{eq:eq:C_UVR}
\end{align}
We see that the sought after correlation  is a correlation coefficient between Gaussian variables, but with $\ld$ samples instead of $T$.  Using again Eq.~\eqref{eq:symmetric_Beta}, we write
\begin{align}\label{eq:Beta_signal}
    \mathrm{pdf}\left((c_{UV})_{pq}\right)=\text{Beta}\left((c_{UV})_{pq};\alpha_\text{s},\alpha_\text{s};-1, 2\right),\, p\neq q,
\end{align}
with parameter
\begin{align}
    \alpha_\text{s}=\frac{\ld-1}{2}\,.
\end{align}
We remind the reader that Eq.~(\ref{eq:Beta_signal}) holds to $\mathcal{O}(T^{-1/2})$. The variance of this density is
\begin{align}
    \text{var}_\text{s}&=\ld^{-1}\,.
\end{align}
This expression agrees with numerical simulations very well, cf.~Fig.~\ref{fig:fig1}.

Finally, for the signal-noise cross terms in the correlation, we have
\begin{align}
    (c_{(UV)^TR})_{pq}&=\frac{1}{T\sigma^\mathrm{s}_p\sigma_q^\mathrm{n}}\sum_t \sum_\mu V_{p\mu}U_{\mu t}R_{tq}\nn\\
    &=\frac{1}{\sigma^\mathrm{s}_p\sigma_q^\mathrm{n}}\sum_\mu V_{p\mu}\left(\frac{1}{T}\sum_t U_{\mu t}R_{tq}\right)\,.\label{eq:C_UVR}
\end{align}
For the quantity in parentheses in Eq.~(\ref{eq:C_UVR}), we define
\begin{align}
    r_{\mu p}\equiv\frac{1}{T}\sum_t U_{\mu t}R_{tq}\,.
\end{align}
This is a covariance between two independent Gaussian random numbers and again follows a rescaled form of the distribution in Eq.~(\ref{eq:pdf_X}) with variance $\sigma_U^2{\sigma_q^\mathrm{n}}^2 T^{-1}$. Since $T$ is large, the distribution approaches a Gaussian and we further define $r_{\mu q}\equiv\sigma_U \sigma_q^\mathrm{n}T^{-1/2}r_{\mu q}'$, such that $r_{\mu q}'$ is a unit Gaussian random variable. Thus we obtain
\begin{multline}
    (c_{(UV)^TR})_{pq}
    =\frac{\sigma_U\ld\sigma_q^\mathrm{n}T^{-1/2}}{\sigma^\mathrm{s}_p\sigma_q^\mathrm{n}}\left(\frac1\ld\sum_\mu V_{p\mu}r'_{\mu q}\right)\\
    =\ld^{1/2}T^{-1/2}\left(\frac1\ld\sum_\mu \frac{V_{p\mu}r'_{\mu q}}{\frac{1}{\ld}\sum_\mu {V_{p\mu}}^2}\right)\,,
\end{multline}
where we have extracted the factor of $\ld$ to highlight that the expression in parenthesis is the correlation between Gaussian random numbers. From this, using Eq.~\eqref{eq:symmetric_Beta}, we conclude that 
\begin{align}\label{eq:Beta_signal_noise}
    \mathrm{pdf}&\left((c_{(UV)^TR})_{pq}\right)=\nn\\
    &\text{Beta}\left((c_{(UV)^TR})_{pq};\alpha_\text{sn},\alpha_\text{sn};-1, 2\right),\, p\neq q\,,
\end{align}
with parameter
\begin{align}
    \alpha_\text{sn}=\frac{\ld^{1/2}T^{1/2}-1}{2}\,.
\end{align}
The variance of this density is
\begin{align}\label{eq:var_sn}
    \text{var}_\text{sn}&=\ld^{-1/2}T^{-1/2}=\sqrt{\text{var}_\text{s}\cdot \text{var}_\text{n}}\,.
\end{align}
An analogous expression holds for the $R^TUV$ contribution.

The empirical correlation matrix is given by
\begin{align}
    c_{pq}=\frac1T\sum_t\frac{X_{pt}X_{tq}}{\sigma^\mathrm{sn}_p\sigma^\mathrm{sn}_q}\,,
\end{align}
where
\begin{align}
    (\sigma^\mathrm{sn}_p)^2=(\sigma^\mathrm{s}_p)^2+(\sigma^\mathrm{n}_p)^2\,.
\end{align}
Using Eqs~\eqref{eq:C_R}, \eqref{eq:C_UV} and \eqref{eq:C_UVR}, the correlation matrix can be written as a weighted sum of the three types of contributions
\begin{align}
    c_{pq}&=\frac{\sigma^\mathrm{s}_p\sigma^\mathrm{s}_q}{\sigma^\mathrm{sn}_p\sigma^\mathrm{sn}_q}(c_{UV})_{pq}+\frac{\sigma^\mathrm{s}_p\sigma^\mathrm{n}_q}{\sigma^\mathrm{sn}_p\sigma^\mathrm{sn}_q}(c_{(UV)^TR})_{pq}\nn\\
    &+\frac{\sigma^\mathrm{n}_p\sigma^\mathrm{s}_q}{\sigma^\mathrm{sn}_p\sigma^\mathrm{sn}_q}(c_{R^TUV})_{pq}+\frac{\sigma^\mathrm{n}_p\sigma^\mathrm{n}_q}{\sigma^\mathrm{sn}_p\sigma^\mathrm{sn}_q}(c_{R})_{pq}
\end{align}
Each term on the right-hand side of this equation follows a Beta distribution as computed above. However, the $\alpha$ parameter of each distribution is modified by the corresponding weight in the above sum. Consequently, the variance of each distribution is rescaled by the weight:
\begin{align}
    \mathrm{var}'_\mathrm{s}&=\frac{\sigma^\mathrm{s}_p\sigma^\mathrm{s}_q}{\sigma^\mathrm{sn}_p\sigma^\mathrm{sn}_q}\mathrm{var}_\mathrm{s}\,\\
    \mathrm{var}'_\mathrm{n}&=\frac{\sigma^\mathrm{n}_p\sigma^\mathrm{n}_q}{\sigma^\mathrm{sn}_p\sigma^\mathrm{sn}_q}\mathrm{var}_\mathrm{n}\,\\
    \mathrm{var}'_\mathrm{sn}&=\frac{\sigma^\mathrm{s}_p\sigma^\mathrm{n}_q}{\sigma^\mathrm{sn}_p\sigma^\mathrm{sn}_q}\mathrm{var}_\mathrm{sn}\,.
\end{align}
To determine an expression for the combined distribution of signal and noise correlations, we make use of the observation that the sum of Beta distributions can be well approximated by a single Beta distribution~\cite{sum_iid_Beta_distributed_variables_app}. We determine the parameters of the Beta distribution by adding the means and variances of the distributions in the sum and analytically match the parameter of the single Beta distribution. 

The means of the Beta distributions in Eq.~\eqref{eq:Beta_noise}, Eq.~\eqref{eq:Beta_signal}, and Eq.~\eqref{eq:Beta_signal_noise} are zero and thus the mean of the density of the combined contributions is also zero.  Taking the sum of variances we obtain
\begin{align}\label{eq:var_Beta}
    \mathrm{var}=\mathrm{var}'_\mathrm{s}+\mathrm{var}'_\mathrm{n}+\mathrm{var}'_\mathrm{sn}+\mathrm{var}'_\mathrm{ns}\,.
\end{align}
In the limit when $T$ and $\ld$ are large enough such that contributions of $\mathcal O(T^{-1/2})$ and $\mathcal O(\ld^{-1/2})$ can be neglected, we have the following convergence of the empirical quantities
\begin{align}
    (\sigma_p^\mathrm{s})^2&\rightarrow \ld\sigma_U^2\sigma_V^2\,,\\
    (\sigma_p^\mathrm{n})^2&\rightarrow \sigma^2\,,\\
    (\sigma_p^\mathrm{sn})^2\,,(\sigma_p^\mathrm{ns})^2&\rightarrow \ld\sigma_U^2\sigma_V^2+\sigma^2\,.
\end{align}
Consequently the variances of the contributions take the form
\begin{align}
    \mathrm{var}'_\mathrm{s}&\rightarrow\frac{m^{-1}}{1+\textsf{SNR}^{-1}}\,,\\
    \mathrm{var}'_\mathrm{n}&\rightarrow\frac{T^{-1}}{1+\textsf{SNR}}\,,\\
    \mathrm{var}'_\mathrm{sn}\,,\mathrm{var}'_\mathrm{ns} &\rightarrow\frac{\ld^{-1/2} T^{-1/2}}{\sqrt{1+\textsf{SNR}}\sqrt{1+\textsf{SNR}^{-1}}}\,,\\
\end{align}
Thus, in this limit, the variance of the Beta distribution, Eq.~\eqref{eq:var_Beta}, is of the form
\begin{align}
    \text{var}&\approx\left(\frac{\ld^{-1/2}}{\sqrt{1+\textsf{SNR}^{-1}}}+\frac{T^{-1/2}}{\sqrt{1+\textsf{SNR}}}\right)^2\,.
\end{align}
Finally, from the relation in Eq.~\eqref{eq:var_alpha_relation}, we obtain the parameter $\alpha$ of the sought after Beta distribution. 
\begin{align}
    \alpha=\frac{\text{var}^{-1}-1}{2}\,.
\end{align}
A comparison between the analytic form of the density and simulated data is shown in Fig.~\ref{fig:fig1} for $\mathsf{SNR}\rightarrow\infty$, and in Fig.~\ref{fig:correlations_finitesmallSNR} for finite $\mathsf{SNR}$ and $\mathsf{SNR}\rightarrow0$. In the extreme noise limits, the analytic form closely matches the simulation. In the large noise limit of $\mathsf{SNR}\rightarrow0$, shown in Fig.~\ref{fig:correlations_finitesmallSNR} (b), the density is close to a Gaussian, because the number of observations $T$ is large. In the regime of finite $\mathsf{SNR}$, shown in Fig.~\ref{fig:correlations_finitesmallSNR} (a), deviations between the analytic form and the simulation appear for small values of $\ld$. We expect that these deviations will disappear by removing the various approximations made in the above analytic derivation.

\section{Spectrum of the normalized empirical covariance Matrix}\label{app:eigenvalue_density}

To compute the eigenvalue density of the NECM $\mathbf C$, we use methods of  Random Matrix Theory~\cite{Potters_Bouchaud_2020_App}. The standard approach is to compute the finite size Stieljtes transform
\begin{align}
g^N_{\mathbf C}(z)=\frac1N\text{Tr}(z\mathbf I-\mathbf C)^{-1}\,,
\end{align}
where $\mathbf I$ is the identity matrix, $z\in\mathbb C$ and $g^N_{\mathbf C}$ is a complex function. In the limit of large matrices -- large $N$ or thermodynamic limit -- the finite size Stieltjes transform becomes, $\mathfrak g_\mathbf{C} (z)$. Then the eigenvalue density is obtained as the imaginary part of the limit of the Stieltjes transform: 
\begin{align}\label{eq:sokhotski-plemelj_app}
    \rho(\lambda)=\frac{1}{\pi}\lim_{\eta\rightarrow 0^+}\Im \mathfrak g(z=\lambda-i\eta)\,,
\end{align}
where $\Im$ denotes the imaginary part.

We start with writing again the definition  of the normalized empirical covariance  matrix (NECM), which differs from the correlation matrix only by  $\mathcal{O}(T^{-1/2})$:
\begin{align}\label{eq:C_full}
\mathbf C=&\frac1T\widetilde{\mathbf X}^T\widetilde{\mathbf X}=\frac1T(\widetilde{\mathbf U\mathbf V+\sigma\mathbf R})^T(\widetilde{\mathbf U\mathbf V+\sigma \mathbf R})\nn\\
    =&\frac{1}{T}\Big((\widetilde{\mathbf U\mathbf V})^T(\widetilde{\mathbf U\mathbf V})+\tilde\sigma^2\mathbf R^T\mathbf R\nn\\
    &+\tilde\sigma(\widetilde{\mathbf U\mathbf V})^T\mathbf R+\tilde \sigma\mathbf R^T\widetilde{\mathbf U\mathbf V}\Big)\,.
\end{align}
The NECM contains three different contributions: the $(\mathbf{UV})^T(\mathbf{UV})$ from the pure latent feature signal, $\mathbf R^T\mathbf R$ from pure noise, and two terms of the type $(\mathbf{UV})^T\mathbf R$, which are cross terms between the latent signal and the noise. Each contribution is an $N\times N$ random matrix. Critical to computing the eigenvalue density of random matrices is the concept of \emph{matrix freeness}~\cite{voiculescu1992free_App}, which is the generalization of statistical independence to matrices. The eigenvalue spectrum of sums and products of free matrices can be computed from spectra of summands and factors using the $\mathcal R$- and the $\mathcal S$-transforms, which are related to the Stieltjes transform $\mathfrak g$ and are additive and multiplicative, respectively. The signal-signal and the noise-noise contributions in the NECM definition are certainly free w.\ r.\ t.\ each other.  We will argue in \emph{Appendix \ref{app:signal_noise_approx}}  that, in our regimes of interest (the zero-noise limit ($\mathsf{SNR}\rightarrow\infty$), the classical statistics limit from Eq.~\eqref{eq:classic_limit}, and intensive limit from Eq.~\eqref{eq:intensive}), the cross-term contributions are negligible, so that we can drop them and approximate the NECM as
\begin{align}\label{eq:EC_approx}
    \mathbf C&\approx\frac{(\mathbf U\mathbf V)^T(\mathbf U\mathbf V)+\sigma^2\mathbf R^T\mathbf R}{\sigma_X^2T}:=\mathbf C_{\widetilde{\mathbf U\mathbf V}}+\mathbf C_{\widetilde\sigma \mathbf R}\,,
\end{align}
so that free matrix theory applies. 

\subsection{Parameterizing the random matrix problem and the large matrix limit}\label{app:parameterisation}
To calculate the spectrum of the signal-signal contribution to the NECM, 
\begin{align}
    \mathbf C_{\widetilde{\mathbf{UV}}}=\frac{1}{\sigma_X^2T}(\mathbf U\mathbf V)^T(\mathbf U\mathbf V)\,,
\end{align}
we note that, assuming $\ld<T,N$, this $N\times N$ matrix is of rank $\ld$. Thus we can work in the basis, where
\begin{align}
    \mathbf C_{\widetilde{\mathbf{UV}}}=\left(
\begin{array}{cc} 
\mathbf H_{\widetilde{\mathbf{UV}}} & 0\\ 
0 & 0
\end{array}
    \right)\,,
\end{align}
and
\begin{align}\label{eq:Cprime_UV}
\mathbf H_{\widetilde{\mathbf{UV}}}=\frac{1}{\sigma_{X}^2T}\left(\mathbf U^T\mathbf U\right)\left(\mathbf V\mathbf V^T\right)\,.
\end{align}
There are $\ld$ non-trivial eigenvalues associated with $\mathbf H$, while the remaining $N-\ld$ eigenvalues are zero.
The finite size Stieltjes transform, $g_{\mathbf C}^N=N^{-1}\text{Tr}(z\mathbf I-{\mathbf C}_{\widetilde{\mathbf{UV}}})^{-1}$, is then of the form
\begin{align}\label{eq:finiteN_Stieltjes}
    g^N_{\mathbf C_{\widetilde{\mathbf{UV}}}}(z)&=\frac1N\left(\ld\,\frac{1}{\ld}\sum_{\mu=1}^\ld\frac{1}{z-\lambda_\mu}+\frac{N-m}{z}\right)\nn\\
    &=\frac1N\left(\ld\,h^m_{\mathbf H_{\widetilde{\mathbf{UV}}}}(z)+\frac{N-m}{z}\right)\,,
\end{align}
where $\lambda_\mu$ are the $\ld$ eigenvalues of $\mathbf H_{\widetilde{\mathbf{UV}}}$ and $h^m_{\mathbf H_{\widetilde{\mathbf{UV}}}}(z)$ is its finite size Stieltjes transform. 

Now we note that $\mathbf H_{\widetilde{\mathbf{UV}}}$ in Eq.~\eqref{eq:Cprime_UV} is the product of two white Wishart matrices
\begin{align}\label{eq:H_UV_N}
\mathbf H_{\widetilde{\mathbf{UV}}}=\frac{N}{\sigma_{X}^2}\,\mathbf W_{\mathbf U}\mathbf W_{\mathbf V^T}\,,
\end{align}
where 
\begin{align}\label{eq:white_Wishart}
    \mathbf W_{\mathbf Y}=\frac{1}{T}\mathbf Y^T\mathbf Y\,,
\end{align}
is the Wishart matrix, and $\mathbf Y$ is a $T\times N$ matrix with i.i.d.\ standard normal entries. The key parameter characterizing such standard $\mathbf W_{\mathbf Y}$ is the ratio of the number of columns to that of  rows
\begin{align}
    q\equiv\frac NT\,.
\end{align}
Since $\mathbf U$ and $\mathbf V^T$ are $T\times \ld$ and $N\times \ld$  matrices, respectively, a natural characterisation of $\mathbf H_{\widetilde{\mathbf U\mathbf V}}$ is then
\begin{align}\label{eq:q_defns}
    q\equiv\frac{N}{T},\quad q_U\equiv\frac{\ld}{T},\quad q_{V^T}\equiv\frac{\ld}{N}\,, 
\end{align}
with  $q\,q_{V^T}=q_U$, so that there are only two independent parameters. 

It is now convenient to define
\begin{align}
    \sigma_X^2=\ld\left(\sigma_U^2\sigma_V^2+\frac{\sigma^2}{\ld}\right)\equiv\ld\bar\sigma_X^2\,,
\end{align}
where we used Eq.~\eqref{eq:UV_variance}, so that Eq.~\eqref{eq:H_UV_N} becomes 
\begin{align}\label{eq:H_UV}
\mathbf H_{\widetilde{\mathbf{UV}}}=\frac{1}{q_{V^T}\bar\sigma_{X}^2}\,\mathbf W_{\mathbf U}\mathbf W_{\mathbf V^T}\,.
\end{align}

In the following,  we only consider the limit of large matrices. Here  $T$, $N$, $\ld$ and $\sigma^2$ go to infinity in such a way that $q$, $q_{V^T}$ and $\textsf{SNR}$ are all constant. Then in the thermodynamic limit the finite size Stieltjes transform in Eq.~\eqref{eq:finiteN_Stieltjes} becomes
\begin{align}\label{eqn:C_UV_Stieltjes}
    \mathfrak g_\mathbf{C_{\widetilde{\mathbf{UV}}}}
    &=q_{V^T}\mathfrak h+\frac{1-q_{V^T}}{z}\,,
\end{align}
where $\mathfrak g_\mathbf{C_{\widetilde{\mathbf{UV}}}}$ and $\mathfrak h$ are the large matrices limits of the Stieltjes transforms of  $g^N_{\mathbf C_{\widetilde{\mathbf{UV}}}}$ and $h^m_{\mathbf H_{\widetilde{\mathbf{UV}}}}$, respectively.

\subsection{The spectrum of $\mathbf{C}_{\widetilde{\mathbf{UV}}}$}\label{app:C_UV_spectrum}
We now compute the eigenvalue density of $\mathbf C_{\widetilde{\mathbf{UV}}}$. The first step is to compute the Stieljtes transform $\mathfrak h$. From Eq.~\eqref{eq:H_UV}, it is clear that this reduces to the problem of computing the eigenvalue spectrum of a product of two Wishart matrices. 

The spectrum of a product of two free matrices can be computed with the help of the $\mathcal S$-transform, which is defined for a random matrix  $\mathbf A$ as
\begin{align}\label{eq:S_transform}
    \mathcal S_\mathbf{A}(t)=\frac{t+1}{t{\mathcal T}_\mathbf{A}^{-1}(t)}\,,
\end{align}
where ${\mathcal T}_\mathbf{A}^{-1}(t)$ is the functional inverse of the $\mathcal T$-transform $\mathcal T_\mathbf{A}(z)$. In turn, the $\mathcal T$-transform is related to the Stieltjes transform of $\mathbf A$ through the relation
\begin{align}\label{eq:T-transform_defn}
\mathcal T_\mathbf{A}(z)=z\mathfrak g_\mathbf{A} (z)-1\,.
\end{align}
Crucially, for free matrices $\mathbf A$ and $\mathbf B$, the $\mathcal S$-transform is multiplicative
\begin{align}\label{eq:S_multiplicative}
    \mathcal S_{\mathbf A\mathbf B}(t)=\mathcal S_{\mathbf A}(t)\mathcal S_{\mathbf B}(t)\,,
\end{align}
and, furthermore, for a scalar $a$,
\begin{align}\label{eq:S_scaling}
    \mathcal S_{a \mathbf A}(t)=a^{-1}\mathcal S_{\mathbf A}(t)\,.
\end{align}

For the white Wishart matrix,  Eq.~\eqref{eq:white_Wishart}, the $\mathcal S$-transform is known to be~\cite{Potters_Bouchaud_2020_App}
\begin{align}\label{eq:Stransf_Wishart}
    \mathcal S_{\mathbf{W_Y}} (t)=\frac{1}{1+qt}\,.
\end{align}
Thus we only need to use the multiplicative property of the $\mathcal S$-transform to compute the signal-signal contributions to the NECM. Specifically,
\begin{align}
    \mathcal S_{\mathbf H_{\widetilde{\mathbf{UV}}}}(t)&=q_{V^T}\bar\sigma_{X}^2\mathcal S_{\mathbf W_{\mathbf U}}\mathcal S_{\mathbf W_{\mathbf V^T}}\nn\\
    &=\frac{q_{V^T}\bar\sigma_{X}^2}{(1+q_Ut)(1+q_{V^T}t)}\,.
\end{align} 
Equation~\eqref{eq:S_transform} then yields 
\begin{align}
    {\mathcal T}^{-1}_{\mathbf H_{\widetilde{\mathbf{UV}}}}(t)=\frac{t+1}{t\,\mathcal S_{\mathbf H_{\widetilde{\mathbf{UV}}}}(t)}
    =\frac{t+1}{t}\frac{(1+q_Ut)(1+q_{V^T}t)}{q_{V^T}\bar\sigma_{X}^2}\,.
\end{align}
We now solve the equation for the functional inverse, ${\mathcal T}^{-1}(\mathcal T(z))=z$, using the definition of the $\mathcal T$-transform, Eq.~\eqref{eq:T-transform_defn}, and dividing by a common factor of $z$. We obtain a cubic equation for the Stieltjes transform $\mathfrak h$:
\begin{align}
   \mathfrak h^3&z^2q_Uq_{V^T}+ \mathfrak h^2 z\left(q_{V^T}(1-q_U)+q_U(1-q_{V^T})\right)\nn\\
   &+\mathfrak h \left((1-q_U)(1-q_{V^T})-z q_{V^T}\bar\sigma_X^2\right)
   + q_{V^T}\bar\sigma_X^2=0\,.
\end{align}
Finally, we divide by $q_{V^T}\bar\sigma_X^2$ to obtain 
\begin{align}\label{eq:polynomial_lowrank}
   \mathfrak h^3\frac{z^2q_U}{\bar\sigma_X^2}&+ \mathfrak h^2 \frac{z}{\bar\sigma_X^2}\left(1+q-2q_U\right)\nn\\
    &+\mathfrak h \left(\frac{q_{V^T}^{-1}-q-1+q_U}{\bar\sigma_X^2}-z\right)+ 1=0\,.
\end{align}
Similar equations for the Stieljtes transform of the product of two random matrices have been stated in~\cite{muller2002random,PhysRevE.82.061114_app,dupic2014spectral_app}. Their polynomials differ from Eq.~\eqref{eq:polynomial_lowrank} in details, because we consider the Stieljtes transform of the covariance matrix including a theoretical normalisation factor.

The next step is to solve Eq.~\eqref{eq:polynomial_lowrank} analytically in the classical statistics limit and the intensive limit. We remind the reader that, for the pure signal contribution, we work in the zero noise limit $\textsf{SNR}\rightarrow\infty$, such that
\begin{align}
    \bar\sigma_X^2\equiv\frac{\sigma_X^2}{\ld}=\sigma_U^2\sigma_V^2(1+\textsf{SNR}^{-1})=\sigma_U^2\sigma_V^2\,.
\end{align}

\subsubsection{Classical statistics limit}
In the  \textit{classical statistics} limit, Eq.~\eqref{eq:classic_limit}, the polynomial equation for the Stieltjes transform, Eq.~\eqref{eq:polynomial_lowrank}, becomes:
\begin{align}
    \mathfrak h^2 \frac{z}{\bar\sigma_X^2}+\mathfrak h \left(\frac{q_{V^T}^{-1}-1}{\bar\sigma_X^2}-z\right)+1=0\,.
\end{align}
The discriminant is 
\begin{align}
    \Delta=z^2-2\frac{1+q_{V^T}^{-1}}{\bar\sigma_X^2}z+\left(\frac{q_{V^T}^{-1}-1}{\bar\sigma_X^2}\right)^2,
\end{align}
and the roots of the discriminant are
\begin{align}\label{eq:classical_bound_SNRinf_app}
    \lambda_{\pm}^{\infty}=
        \bar\sigma_X^{-2}\left(1\pm \sqrt{q_{V^T}}^{-1}\right)^2.
\end{align}
We thus obtain
\begin{align}\label{eq:hm_classcial}
    \mathfrak{h}_\pm=\frac{-\frac{q_{V^T}^{-1}-1}{\bar\sigma_X^2}+z\pm\sqrt{(z-\lambda_-^{\infty})(z-\lambda_+^{\infty})}}{2 z{\bar\sigma_X}^{-2}}\,.
\end{align}
To obtain $\mathfrak g_{\mathbf C_{\widetilde{\mathbf{UV}}}}$, we now need to add the contribution of the zero eigenvalues:
\begin{align}\label{eq:g_C_UV}
    \mathfrak g_{\mathbf C_{\widetilde{\mathbf{UV}}}}&=q_{V^T}\mathfrak h_\pm+\frac{1-q_{V^T}}{z}
    \nn\\
    &=\frac{1-q_{V^T}}{2z}+\frac{q_{V^T}}{2 {\bar\sigma_X}^{-2}}\pm\frac{\sqrt{(z-\lambda_-^{\infty})(z-\lambda_+^{\infty})}}{2 zq_{V^T}^{-1}{\bar\sigma_X}^{-2}}\,.
\end{align}

We are now ready to obtain the eigenvalue density, as in Eq.~\eqref{eq:sokhotski-plemelj_app}. While this is a standard calculation~\cite{Livan_2018_app}, we summarise it here for the reader's benefit. The second term on the right-hand side of Eq.~\eqref{eq:g_C_UV} is real, does not contribute to the imaginary part, and we ignore it. For the first and the third terms, we multiply the numerators and the denominators by $z^*=\lambda+i\eta$. The imaginary part of the first term is then
\begin{align}\label{eq:classical_density_oneoverz}
    \Im \left(\frac{1-q_{V^T}}{2z}\right)=\frac{(1-q_{V^T})\eta}{2(\eta^2+\lambda^2)}=\frac{(1-q_{V^T})\pi}{2}\delta_\eta(\lambda)\,,
\end{align}
where we have used the definition of the Lorentz curve, $\delta_\eta(\lambda)=\pi^{-1}\eta/(\eta^2+\lambda^2)$.
For the third term, the crucial step is to rewrite the square root using the relation
\begin{align}
    \sqrt{a+ib}=P+iQ\,,
\end{align}
where $a$ and $b$ are real, $b\neq0$ and
\begin{align}
    P&=\frac{1}{\sqrt{2}}\sqrt{\sqrt{a^2+b^2}+a}\nn,\\
    Q&=\frac{\text{sgn}(b)}{\sqrt{2}}\sqrt{\sqrt{a^2+b^2}-a}\,,
\end{align}
where $\mathrm{sgn}(x)=1$ for $x>0$ and $-1$ for $x<0$~\cite{Rabinowitz_1993_app}. For the argument of the square root in the third term of Eq.~\eqref{eq:g_C_UV}, we find
\begin{align}
    a&=\lambda^2 - \eta^2 + \lambda_+^\infty\lambda_-^\infty - (\lambda_+^\infty+\lambda_-^\infty) \lambda\nn,\\
    b&=(-2\lambda + \lambda_+^\infty+\lambda_-^\infty)\eta\,.
\end{align}
The imaginary part of the third term takes the form
\begin{align}\label{eq:density_classical_sqrt}
    &\Im{\left(\pm\frac{\sqrt{(z-\lambda_-^{\infty})(z-\lambda_+^{\infty})}}{2 zq_{V^T}^{-1}{\bar\sigma_X}^{-2}}\right)}=\nn\\
    &=\pm\frac{\Im \left(z^*\left[P+iQ\right]\right)}{2 q_{V^T}^{-1}{\bar\sigma_X}^{-2}|z|^2}\nn\\
    &=\pm\frac{1}{2 q_{V^T}^{-1}{\bar\sigma_X}^{-2}}\left(\frac{\eta}{\eta^2+\lambda^2}P+\frac{\lambda}{\eta^2+\lambda^2}Q\right)\nn\\
    &=\pm\frac{1}{2 q_{V^T}^{-1}{\bar\sigma_X}^{-2}}\left(\pi\delta_\eta(\lambda)P+\frac{\lambda}{\eta^2+\lambda^2}Q\right).
\end{align}

The final step to evaluate Eq.~\eqref{eq:sokhotski-plemelj_app} and to obtain the eigenvalue density, is to take the limit $\eta\rightarrow0^+$. In this limit, the Lorentz curve in Eq.~\eqref{eq:classical_density_oneoverz} converges to the Dirac $\delta$-function. Combining Eqs.~(\ref{eq:classical_density_oneoverz},~\ref{eq:density_classical_sqrt}) yields
\begin{align}\label{eq:Stieltjes_to_rho_infty}
        \rho^\infty(\lambda)&=\frac{1}{\pi}\lim_{\eta\rightarrow0^+}\Im\,\mathfrak g_{\mathbf C_{\widetilde{\mathbf{UV}}}}\nn\\
        &=\pm\frac{\lim_{\eta\rightarrow0^+}P}{2q_{V^T}^{-1}\bar\sigma_X^{-2}}\delta(\lambda)\pm\frac{\lim_{\eta\rightarrow0^+}Q}{2\pi\lambda\bar\sigma_X^{-2}q_{V^T}^{-1}}+\frac{1-q_{V^T}}{2}\delta(\lambda)
\end{align}
with
\begin{align}
    \lim_{\eta\rightarrow0^+}P=\sqrt{\lambda_+^\infty\lambda_-^\infty}=\bar\sigma_X^{-2}(1-q_{V^T}^{-1})\,,
\end{align}
where we have used the expression for the zero noise eigenvalue bounds in Eq.~\eqref{eq:classical_bound_SNRinf_app}, and
\begin{align}
    \lim_{\eta\rightarrow0^+}Q&=\frac{\text{sgn}(b)}{\sqrt{2}}\sqrt{2|(\lambda-\lambda_-^\infty)(\lambda_+^\infty-\lambda)|}\nn\\
    &=\text{sgn}(b)\sqrt{(\lambda-\lambda_-^\infty)(\lambda_+^\infty-\lambda)},
\end{align} 
when $\lambda\in[\lambda_-^\infty,\lambda_+^\infty]$, and the expression vanishes elsewhere.
The $\pm$ signs in Eq.~\eqref{eq:Stieltjes_to_rho_infty} are chosen such as to obtain a physically meaningful eigenvalue density.
Finally, we find the following form of the eigenvalue density
\begin{align}
    \rho^{\infty}(\lambda)&=\frac{\sqrt{(\lambda-\lambda_-^{\infty})(\lambda_+^{\infty}-\lambda)}}{2 \pi \lambda{\bar\sigma_X}^{-2}q_{V^T}^{-1}}+(1-q_{V^T})\delta(\lambda)\,,
\end{align}
with $\bar\sigma_X^2\equiv\sigma_X^2/\ld=\sigma_U^2\sigma_V^2$. We note that in~\cite{cui2020perturbative_app} an expression for an eigenvalue density was given in the special case when $T=N=\ld$ and not including our theoretical normalisation factor.

\subsubsection{Intensive limit}
For the \textit{intensive limit}, Eq.~\eqref{eq:intensive}, the polynomial equation for the Stieltjes transform, Eq.~\eqref{eq:polynomial_lowrank}, becomes
\begin{align}
    \mathfrak h^2 \frac{z}{\bar\sigma_X^2}\left(1+q\right)+\mathfrak h \left(\frac{q_{V^T}^{-1}-q-1}{\bar\sigma_X^2}-z\right)+1=0\,.
\end{align}
The discriminant is 
\begin{align}
    \Delta=z^2-2\frac{q_{V^T}^{-1}+q+1}{\bar\sigma_X^2}z+\left(\frac{q_{V^T}^{-1}-q-1}{\bar\sigma_X^2}\right)^2\,.
\end{align}
The roots of the discriminant are
\begin{align}
    \lambda_{\pm}^{\infty}
    =\bar\sigma_X^{-2}\left( \sqrt{1+q}\pm \sqrt{q_{V^T}}^{-1}\right)^2\,.
\end{align}
Then the solution of the quadratic equation is
\begin{align}
    \mathfrak{h}_\pm=\frac{-\frac{q_{V^T}^{-1}-q-1}{\bar\sigma_X^2}+z\pm\sqrt{(z-\lambda_-^{\infty})(z-\lambda_+^{\infty})}}{2 z{\bar\sigma_X}^{-2}(1+q)}.
\end{align}
Following a calculation analogous to the classical limit, we now add the contribution of the zero eigenvalues and then  determine the density of the eigenvalues. We find:
\begin{align}\label{eq:intensive_SNRinf_density}
    \rho^{\infty}(\lambda)&=\frac{\sqrt{(\lambda-\lambda_-^{\infty})(\lambda_+^{\infty}-\lambda)}}{2 \pi \lambda{\bar\sigma_X}^{-2}(1+q)q_{V^T}^{-1}}+(1-q_{V^T})\delta(\lambda)\,,
\end{align}
with $\bar\sigma_X^2\equiv\sigma_X^2/\ld=\sigma_U^2\sigma_V^2$.

\subsection{Approximation to neglect the signal-noise cross terms}\label{app:signal_noise_approx}
Now we explore when the contribution of the signal-noise cross terms to the NECM can be neglected. Specifically, we will show that it can be done if  $q_U\rightarrow0$ (that is, the number of measurements is much larger than the number of latent features), which we always assume. To show this,  we compute the eigenvalue bounds, $\lambda_\pm^\text{signal-noise}$, of the signal-noise contribution and compare their scaling with $T$ to the scaling of the pure signal and the pure noise eigenvalue bounds.

For the pure signal contribution, the previous section shows that the eigenvalue bounds $\lambda^\infty_\pm$ are $\bar\sigma_X^{-2}\sim\mathcal{O}(T^0)$. The pure noise eigenvalue bounds, given by the Mar\v cenko-Pastur bounds, scale as
\begin{align}
    \lambda_\pm^\mathrm{MP}\sim 1\pm T^{-1/2}\,,
\end{align}
where $1$ is due to self correlations. On the other hand, the signal-noise cross terms do not have self-correlations, and thus we expect their bounds to scale as 
\begin{align}\label{eq:signal_noise_density_scaling}
    \lambda^\mathrm{signal-noise}_\pm\sim T^{-1/2}\,,
\end{align}
becoming negligible for $T\to\infty$. In {\em Appendix~\ref{app:signal-noise_scaling}}, we show this analytically in the classical statistics limit.  We have not been able to achieve similar results more generally. However, since $q_U\rightarrow0$ also in the intensive limit, we expect similar results to hold there too. To show this, we resort to  numerical simulations.

\begin{figure*}
\includegraphics[width=0.8\linewidth]{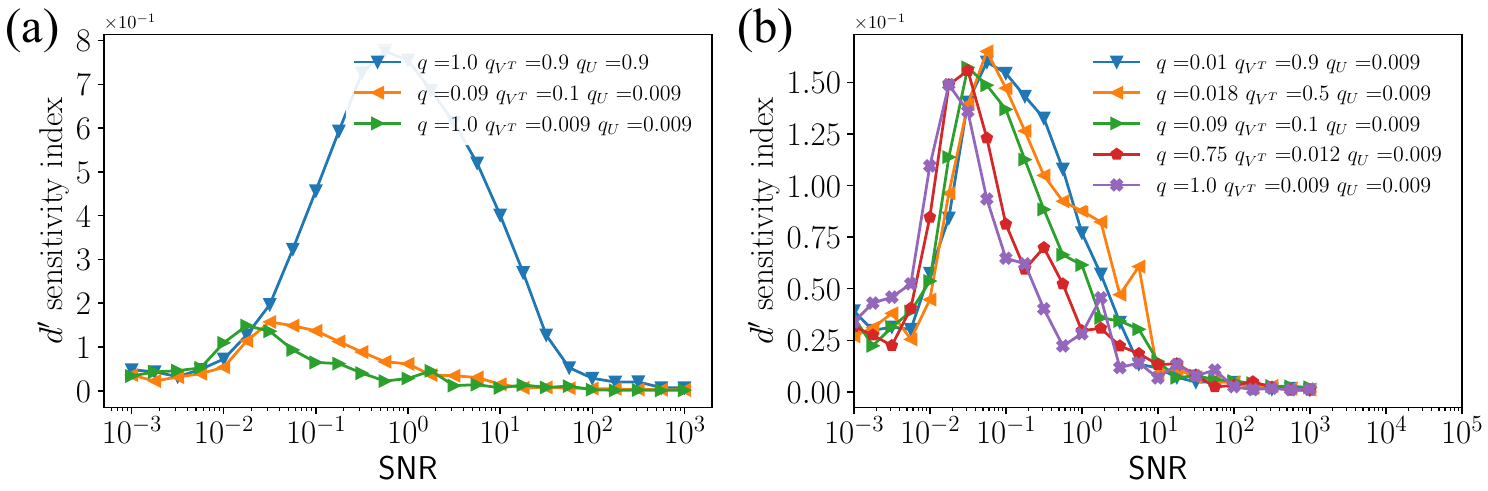}
\caption{Difference between the eigenvalue density of the NECM spectrum with and without the signal-noise cross terms quantified by $d'$. (a) Classical limit (orange), intensive limit (green) and neither of the two limits (blue). (b) Magnified view of $d'$ in the limits of interest:  classical limit (blue, orange and green) and intensive limit (red and purple). Eigenvalue densities are computed from $120$ realisations of the random matrix model. \label{fig:deeprime_vs_SNR}}
\end{figure*}

Specifically, we numerically estimate the Jensen-Shannon divergence between the numerically evaluated eigenvalue densities of the NECM, computed with and without the signal-noise cross terms. To obtain a perceptually intuitive measure of the difference between these distributions, we convert the Jensen-Shannon divergence to the effective sensitivity index $d'$ -- the distance between the means of two unit variance normal distribution with the same Jensen-Shannon divergence as the two analyzed eigenvalue spectra. We investigate the dependence of $d'$ on various choices of our model parameters. The comparison between the spectra of the full and the approximate NECM is shown in Fig.~\ref{fig:deeprime_vs_SNR}. We observe that the sensitivity index reaches a maximum for $\mathsf{SNR}\sim 10^{-1}$, and falls off in the limits of small or  large $\mathsf{SNR}$, where the noise or the signal dominate, respectively. Crucially,  the maximum value of $d'$ is small for $q_U\to0$. Thus neglecting the cross-term contributions to the NECM spectrum in our limits of interest is warranted.

\subsubsection{Scaling behavior of the signal-noise eigenvalue bounds in the classical limit}\label{app:signal-noise_scaling}
We now derive the scaling of the signal-noise eigenvalue spectrum bounds in Eq.~\eqref{eq:signal_noise_density_scaling} in the classical statistics limit. From the NECM in Eq.~(\ref{eq:C_full}), the signal-noise cross terms are of the form
\begin{align}
    \mathbf M\equiv\frac{\tilde\sigma}{T}(\widetilde{\mathbf U\mathbf V})^T{\mathbf{R}}\,.
\end{align}
To compute the spectrum of this matrix, we use the following trick. The singular values of $\mathbf M$ are equal to the square roots of the non-zero eigenvalues of its  square,
\begin{align}
    \mathbf M^2&=\mathbf M\mathbf M^T=\frac{\tilde\sigma^2}{T^2}(\widetilde{\mathbf U\mathbf V})^T{\mathbf{R}}{\mathbf R}^T\widetilde{\mathbf U\mathbf V}\,.
\end{align}
In turn, the non-zero eigenvalues of this $N\times N$ matrix, are  equal to the eigenvalues of the $T\times T$ matrix
\begin{align}\label{eq:hatM}
    \widehat{\mathbf{M}}^2\equiv q^2\tilde\sigma^2\frac{{\mathbf R}{\mathbf R^T}}{N}\frac{(\widetilde{\mathbf U\mathbf V})(\widetilde{\mathbf U\mathbf V})^T}{N}\,.
\end{align}
We note that the r.h.s.\ of the above equation is a product of the $T\times T$ dual correlation matrices $\mathbf C_{\tilde\sigma\mathbf R^T}$ and $\mathbf C_{(\widetilde{\mathbf{UV}})^T}$. To compute the spectrum of the product, we employ the $\mathcal S$-transform formalism, as explained above. The first step is to obtain the $\mathcal S$-transforms of the dual correlation matrices, which we compute from the Stieltjes transform ~\cite{Potters_Bouchaud_2020_App}. For the noise part, we have
\begin{align}\label{eq:noise_dual}
    \mathfrak g_{\mathbf{C}_{\mathbf{R}^T}}(z)=q^2\mathfrak g_{\mathbf{C}_{\mathbf{R}}}(qz)+\frac{1-q}{z}.
\end{align}
For the signal part, we have
\begin{align}\label{eq:signal_dual}
    \mathfrak g_{\mathbf{C}_{(\widetilde{\mathbf{UV}})^T}}(z)=q^2\mathfrak g_{\mathbf{C}_{\widetilde{\mathbf{UV}}}}(qz)+\frac{1-q}{z}\,.
\end{align}
For the noise Wishart matrix, the Stieljtes transform is (cf.~Eq.~\eqref{eq:Stransf_Wishart}):
\begin{align}
    \mathcal S_{\mathbf C_{{{\mathbf R}}^T}}=\frac{1}{1+q^{-1}t}\,.
\end{align}
Including the renormalized noise strength, $\tilde\sigma$, and the additional factor of $q$ from Eq.~\eqref{eq:hatM}, by using the scaling relation Eq.~\eqref{eq:S_scaling}, the $\mathcal S$-transform is
\begin{align}
    \mathcal S_{\mathbf C_{{{q\tilde\sigma\mathbf R}}^T}}=\frac{q^{-2}\tilde\sigma^{-2}}{1+q^{-1}t}\,.
\end{align}
Next, we write the $\mathcal S$-transform of the pure signal part. Evaluating Eq.~\eqref{eq:signal_dual} using Eq.~\eqref{eq:g_C_UV}, we find
\begin{multline}
    \mathfrak g_{\mathbf C_{(\widetilde{\mathbf{UV}})^T}}=\\ \frac{\frac{2-q-qq_{V^T}}{\bar\sigma_X^2}+q^2q_{V^T}z\pm qq_{V^T}\sqrt{(qz-\lambda^\infty_+)(qz-\lambda^\infty_-)}}{2z\bar\sigma_X^{-2}}\,,
\end{multline}
from which we obtain the following equation for the Stieltjes transform:
\begin{align}
    &\left(2z\bar\sigma_X^{-2}\mathfrak g_{\mathbf C_{(\widetilde{\mathbf{UV}})^T}}-\frac{2-q-qq_{V^T}}{\bar\sigma_X^2}-q^2q_{V^T}z\right)^2\nn\\
    &-q^2q_{V^T}^2(qz-\lambda_+^\infty)(qz-\lambda_-^\infty)=0\,.
\end{align}
Using the relation, $\mathcal T=z\mathfrak g-1$, we find the equation for the $\mathcal T$-transform
\begin{multline}
    \mathcal T^2_{\mathbf C_{(\widetilde{\mathbf{UV}})^T}}+\\ \mathcal T_{\mathbf C_{(\widetilde{\mathbf{UV}})^T}}(-zq^2q_{V^T}\bar\sigma_X^2+qq_{V^T}+q)+q^2q_{V^T}=0\,.
\end{multline}
Interpreted as an equation for the functional inverse transform $\mathcal T^{-1}$, this  becomes
\begin{align}
    t^2+t(-\mathcal T^{-1}_{\mathbf C_{(\widetilde{\mathbf{UV}})^T}}q^2q_{V^T}\bar\sigma_X^2+qq_{V^T}+q)+q^2q_{V^T}=0\,.
\end{align}
Now solving for the functional inverse transform, we find
\begin{align}
    \mathcal T^{-1}_{\mathbf C_{(\widetilde{\mathbf{UV}})^T}}=\frac{(t+q)(t+q_{U})}{tqq_U\bar\sigma_X^2},
\end{align}
from which we determine the $\mathcal S$-transform
\begin{align}
    \mathcal S_{\mathbf C_{(\widetilde{\mathbf{UV}})^T}}=\frac{t+1}{t\mathcal T^{-1}_{\mathbf C_{(\widetilde{\mathbf{UV}})^T}}}=\frac{qq_U\bar\sigma_X^2(t+1)}{(t+q)(t+q_U)}\,.
\end{align}

The $\mathcal S$-transform of the product now reads
\begin{align}
    \mathcal S_{\widehat{\mathbf{M}}^2}&=\mathcal S_{\mathbf C_{(\widetilde{\mathbf{UV}})^T}}\mathcal S_{\mathbf C_{{q\tilde\sigma{\mathbf R}}^T}}\nn\\
    &=\frac{t+1}{t}\frac{tqq_U\bar\sigma_X^2}{(t+q)(t+q_U)}\frac{q^{-2}\tilde\sigma^{-2}}{1+q^{-1}t}.
\end{align}
From this we read off the inverse transform
\begin{align}
    \mathcal T^{-1}_{\widehat{\mathbf{M}}^2}&=\frac{(t+q)(t+q_U)(1+q^{-1}t)}{tq^{-1}q_U\bar\sigma_X^2\tilde\sigma^{-2}}\nn\\
    &=\frac{(t+q)(t+q_U)(1+q^{-1}t)}{tq^{-1}q_U\mathsf{SNR}^{-1}}.
\end{align}
The equation for the $\mathcal T$-transform is now
\begin{align}
    (\mathcal T_{\widehat{\mathbf{M}}^2}+q)(\mathcal T_{\widehat{\mathbf{M}}^2}+q_U)(1+q^{-1}\mathcal T_{\widehat{\mathbf{M}}^2})=\dfrac{zq_U\mathcal T_{\widehat{\mathbf{M}}^2}}{q\,\mathsf{SNR}}.
\end{align}
Using $\mathcal T=z\mathfrak g-1$ we write down the cubic polynomial equation for $\mathfrak g$:
\begin{align}
    a\mathfrak g_{\widehat{\mathbf{M}}^2}^3+b\mathfrak g_{\widehat{\mathbf{M}}^2}^2+c\mathfrak g_{\widehat{\mathbf{M}}^2}+d=0\,,
\end{align}
with coefficients
\begin{align}
    a&=z^3,\\
    b&=2 q z^2 + q_U z^2 - 3 z^2,\\
    c&=q^2 z + 2 q q_U z - 4 q z - 2 q_U z + 3 z - \frac{q_U z^2}{\textsf{SNR}},\\
    d&=q^2 q_U - q^2 - 2 q q_U + 2 q + q_U -1 + \frac{q_U z}{\textsf{SNR}}.
\end{align}
The eigenvalue density is non-zero for complex solutions of the equation. The equation admits complex solutions when the discriminant $\Delta$ is negative:
\begin{align}
    \Delta=4P^3+27Q^2\,,
\end{align}
where
\begin{align}
    P&=\frac{3ac-b^2}{3a^2},\\
    Q&=\frac{2b^2-9abc+27a^2d}{27a^3}.
\end{align}
Written out explicitly, the determinant takes the form
\begin{align}
    \Delta=\frac{4(3ac-b^2)^3+(2b^3-9abc+27a^2d)^2}{27a^6}\,.
\end{align}
The equation $\Delta=0$ yields a quadratic equation in $z$, giving the bounds on the eigenvalue density of $\widehat{\mathbf{M}}^2$ 
\begin{align}
    z_\pm= \frac{8 q^2 + 20 q q_U - q_U^2 \pm \sqrt{q_U (8 q + q_U)^3}}{8 q_U\textsf{SNR}^{-1}}.
\end{align}
From the definitions of $q=N/T$ and $q_U=\ld/T$, we see that these bounds scale as
\begin{align}
    z_\pm\sim T^{-1}\,.
\end{align}
Since the singular values of $\mathbf M$ are equal to the square root of the eigenvalues of $\widehat{\mathbf{M}}^2$, the eigenvalue bounds of the signal-noise cross terms thus scale as
\begin{align}
    \lambda^\mathrm{signal-noise}_\pm\sim T^{-1/2}\,.
\end{align}
Thus the contribution of the cross-terms can be neglected in the classical statistics limit. 

\subsection{Adding the noise contribution $\mathbf C_\mathbf{\tilde\sigma\mathbf R}$}

In {\em Appendix~\ref{app:C_UV_spectrum}} we  computed the spectrum of the pure signal contribution $\mathbf C_{\widetilde{\mathbf U\mathbf V}}$ to the NECM in the classical statistics and the intensive limits. Now we will add the pure noise contribution $\mathbf C_{\tilde\sigma\mathbf R}$ to obtain the spectrum of the approximate NECM. Since the noise and the signal contributions are free matrices with respect to each other, the spectrum of their sum can be computed using the $\mathcal R$-transform. The $\mathcal R$-transform of a random matrix $\mathbf A$ is 
\begin{align}\label{eq:Rtransform_defn}
    \mathcal R_\mathbf{A}(z)=\mathcal B_\mathbf{A}(z)-1/z\,,
\end{align}
where the $\mathcal B$-transform is the functional inverse of the Stieltjes transform
\begin{align}
    \mathcal B_{\mathbf A}[\mathfrak g_{\mathbf A}]=z.
\end{align}
The $\mathcal R$-transform is additive for free matrices:
\begin{align}\label{eq:R_transform_additive}
   \mathcal R_{\mathbf A+\mathbf B}(z)=\mathcal R_{\mathbf A}(z)+\mathcal R_{\mathbf B}(z)\,.
\end{align}
It scales according to
\begin{align}
    \mathcal R_{a\mathbf A}(z)=a\mathcal R_{\mathbf A}(az)\,,
\end{align}
where $a$ is a real number. For a white Wishart matrix, Eq.~\eqref{eq:white_Wishart}, the $\mathcal R$-transform is known to be~\cite{Potters_Bouchaud_2020_App}
\begin{align}
    \mathcal R_{\mathbf W_{\mathbf Y}}(z)=\frac{1}{1-qz}\,.
\end{align}
For the pure noise contribution to the NECM, $\mathbf C_{\tilde\sigma\mathbf R}\equiv\tilde\sigma^2\mathbf R^T\mathbf R/T$, this results in
\begin{align}\label{eq:noise_Rtransform}
    \mathcal R_{\mathbf C_{\widetilde\sigma\mathbf R}}(z)=\frac{\widetilde\sigma^2}{1-qz\widetilde\sigma^2}\,.
\end{align}
Our goal is to first obtain the $\mathcal R$-transform of the sum of the signal and the noise contributions
\begin{align}
    \mathcal R_{\mathbf C}(z)=\mathcal R_{\mathbf C_{\widetilde{\mathbf U\mathbf V}}}(z)+\mathcal R_{\mathbf C_{\widetilde\sigma\mathbf R}}(z),
\end{align}
and from this to compute the Stieltjes transform to extract the eigenvalue density. Computing the $\mathcal R$-transform of the pure signal contribution $\mathbf C_{\widetilde{\mathbf{UV}}}$ in the classical statistics and the intensive limit requires additional steps.

\subsubsection{Classical statistics limit}\label{app:classical_limit_noise}
First, we compute the $\mathcal R$-transform of $\mathfrak g_{\mathbf C_{\widetilde{\mathbf{UV}}}}$. This is done by solving the functional inverse equation, $\mathfrak g_{\mathbf C_{\widetilde{\mathbf{UV}}}}[\mathcal B_{\mathbf C_{\widetilde{\mathbf U\mathbf V}}}]=z$, which gives us the $\mathcal B$-transform, from which we compute the $\mathcal R$-transform. From Eq.~\eqref{eq:g_C_UV}, we see that  $\mathcal B$-transform satisfies
\begin{align}
    \mathcal B_{\mathbf C_{\widetilde{\mathbf U\mathbf V}}}\left(\mathcal B_{\mathbf C_{\widetilde{\mathbf U\mathbf V}}}z(-q_{V^T}\bar\sigma_X^2 + z) + q_{V^T}\bar\sigma_X^2 + q_{V^T}z - z\right)=0.
\end{align}
A non-trivial solution of this equation is
\begin{align}
    \mathcal B_{\mathbf C_{\widetilde{\mathbf U\mathbf V}}}=\frac{q_{V^T}\bar\sigma_X^2 + q_{V^T}z - z}{z(q_{V^T}\bar\sigma_X^2 - z)}\,.
\end{align}
Using Eq.~\eqref{eq:Rtransform_defn}, this gives us the $\mathcal R$-transform, $\mathcal R_{\mathbf C_{\widetilde{\mathbf U\mathbf V}}}$. To it we add the $\mathcal R$-transform of the noise, Eq.~\eqref{eq:noise_Rtransform}, and subtract $-1/z$ to get  the $\mathcal B$-transform of the approximate NECM:
\begin{multline}
    \mathcal B_{\mathbf C}=\\ \frac{\widetilde\sigma^2z(q_{V^T}\bar\sigma_X^2 - z) + (-q\widetilde\sigma^2z + 1)(q_{V^T}\bar\sigma_X^2 + q_{V^T}z - z)}{z(q_{V^T}\bar\sigma_X^2 - z)(-q\widetilde\sigma^2z + 1)}\,.
\end{multline}
The final step is to write down and solve the inverse function equation $\mathcal B_{\mathbf C}[\mathfrak g_{\mathbf C}]=z$. This is now equivalent to solving the third order polynomial equation
\begin{align}
    &a\mathfrak g_{\mathbf C}^3+b\mathfrak g_{\mathbf C}^2+c\mathfrak g_{\mathbf C}+d=0\,,\; \mbox{with}\\
    &a=qz\widetilde\sigma^2\,,\\
    &b=-qq_{V^T}z\bar\sigma_X^2\widetilde\sigma^2 + \big((q_{V^T} - 1)q + 1\big)\widetilde\sigma^2 - z\,,\\
    &c=(q - 1)q_{V^T}\bar\sigma_X^2\widetilde\sigma^2 + q_{V^T}z\bar\sigma_X^2 - q_{V^T} + 1\,,\\
    &d=-q_{V^T}\bar\sigma_X^2\,.
\end{align}
Written in terms of the signal-to-noise ratio, $\textsf{SNR}$, the coefficients take the form
\begin{align}\label{eq:classical__eval_wnoise_SNR}
    a&=\frac{qz}{1+\textsf{SNR}}\,,\\
    b&=-\frac{qq_{V^T}z}{\textsf{SNR}} + \frac{(q_{V^T} - 1)q + 1}{1+\textsf{SNR}} - z\,,\label{eq:classical_bcoeff}\\
    c&=\frac{(q - 1)q_{V^T}}{\textsf{SNR}} + q_{V^T}z\big(1+\textsf{SNR}^{-1}\big) - q_{V^T} + 1\,,\\
    d&=-q_{V^T}\big(1+\textsf{SNR}^{-1}\big)\,.
\end{align}
It is possible to solve this cubic equation analytically. However, the expressions become lengthy and provide little insight. Therefore, we rely on the numerical solution of the equation as shown in Fig.~\ref{fig:fig2}, as well as on the following analyses in the limits of small and large noise.

First,  in the limit of the pure signal, $\textsf{SNR}\rightarrow\infty$, we recover  Eq.~\eqref{eq:classical_density_SNRinf}. Similarly, by truncating the polynomial coefficients in the pure noise limit, $\textsf{SNR}\rightarrow0$, at order $\mathcal O(\textsf{SNR}^{-1})$, the MP density is recovered.

We also derive an approximate analytic expression for the bounds of the eigenvalue density, $\lambda_\pm^\textsf{SNR}$, which is valid around both noise limits, $\textsf{SNR}\rightarrow0$ and $\textsf{SNR}\rightarrow\infty$. For this, we approximate the coefficients of the polynomial, noting that the smallest contribution to the coefficients common to both limits comes from terms of order $\mathcal{O}(q/(1+\textsf{SNR}))$ (recall that $q\rightarrow0$ in the classical limit). Neglecting these terms leads to a quadratic polynomial equation for the Stieltjes transform
\begin{align}
    &r\mathfrak g_{\mathbf C}^2+s\mathfrak g_{\mathbf C}+t\approx0\,,\; \mbox{with}\\
    &r=-\frac{qq_{V^T}z}{\textsf{SNR}} +\frac{1}{1+\textsf{SNR}} - z\,,\\
    &s=\frac{(q - 1)q_{V^T}}{\textsf{SNR}} + q_{V^T}z\big(1+\textsf{SNR}^{-1}\big) - q_{V^T} + 1\,,\\
    &t=-q_{V^T}\big(1+\textsf{SNR}^{-1}\big)\,.
\end{align}
The approximate bounds of the eigenvalue density then are given by the roots of the discriminant, $\Delta \mathfrak g_{\mathbf C}\approx s^2-4rt$, which gives the following bounds for the nonzero range of the eigenvalue density:
\begin{widetext}
    \begin{align}
        \lambda_\pm^{\textsf{SNR}}&\approx\frac{1+q_{V^T}^{-1}}{1+\textsf{SNR}^{-1}}+\frac{1+q}{1+\textsf{SNR}}\pm2\sqrt{\frac{q_{V^T}^{-1}}{\left(1+\textsf{SNR}^{-1}\right)^2}+\frac{q}{\left(1+\textsf{SNR}\right)^2}+\frac{q}{\left(\sqrt{\textsf{SNR}}+\sqrt{\textsf{SNR}}^{-1}\right)^2}}\nn\\
        &\approx\frac{1+q_{V^T}^{-1}}{1+\textsf{SNR}^{-1}}+\frac{1+q}{1+\textsf{SNR}}\pm2\sqrt{\frac{q_{V^T}^{-1}}{\left(1+\textsf{SNR}^{-1}\right)^2}+\frac{q}{\left(1+\textsf{SNR}\right)^2}}\nn\\
        &=\frac{1+q_{V^T}^{-1}}{1+\textsf{SNR}^{-1}}+\frac{1+q}{1+\textsf{SNR}}\pm2\sqrt{\left[\sqrt{\frac{q_{V^T}^{-1}}{\left(1+\textsf{SNR}^{-1}\right)^2}}+\sqrt{\frac{q}{\left(1+\textsf{SNR}\right)^2}}\right]^2 -\frac{2\sqrt{qq_{V^T}^{-1}}}{(1+\textsf{SNR}^{-1})(1+\textsf{SNR})}}\nn\\
        &\approx\frac{1+q_{V^T}^{-1}}{1+\textsf{SNR}^{-1}}+\frac{1+q}{1+\textsf{SNR}}\pm2\left[\sqrt{\frac{q_{V^T}^{-1}}{\left(1+\textsf{SNR}^{-1}\right)^2}}+\sqrt{\frac{q}{\left(1+\textsf{SNR}\right)^2}}\right]\nn\\
        &=\frac{1}{1+\textsf{SNR}^{-1}}\lambda_\pm^{\infty}+\frac{1}{1+\textsf{SNR}}\lambda_\pm^\text{MP}\,.
    \end{align}
In the second line, we drop the third term under the square root, since it is small in either of the two noise limits. 
In the third line, we have used $(\sqrt{a}+\sqrt{b})^2=a+b+2\sqrt{ab}$, and in the fourth line, we dropped the last term under the square root since it is also small in either of the two noise limits. In the final line we recognize that the terms form the weighted average of $\lambda^{\infty}_\pm$, Eq.~\eqref{eq:classical_bound_SNRinf},  and the Mar\v cenko-Pastur bounds $\lambda_\pm^\text{MP}$.

\subsubsection{Intensive limit}\label{app:intensive_limit_noise}
First we compute the $\mathcal R$-transform of $\mathfrak g_{\mathbf C_{\widetilde{\mathbf{UV}}}}$. This is obtained by solving the functional inverse equation $\mathfrak g_{\mathbf C_{\widetilde{\mathbf{UV}}}}[\mathcal B_{\mathbf C_{\widetilde{\mathbf{UV}}}}]=z$, which gives us the $\mathcal B$-transform, from which we compute the $\mathcal R$-transform. To do this, we employed the symbolic algebra Python library SymPy v1.6.2. The $\mathcal B$-transform satisfies the quadratic equation
    \begin{multline}
        \mathcal B_{\mathbf C_{\widetilde{\mathbf{UV}}}}^2 z (q^2 z - q q_{V^T} \bar\sigma_X^2 + 2 q z - q_{V^T} \bar\sigma_X^2 + z) \\+ \mathcal B_{\mathbf C_{\widetilde{\mathbf{UV}}}} (q^2 q_{V^T} z - 2 q^2 z + q q_{V^T} \bar\sigma_X^2 + 2 q q_{V^T} z - 3 q z + q_{V^T} \bar\sigma_X^2 + q_{V^T} z - z) - q^2 q_{V^T} + q^2 - q q_{V^T} + q=0\,,
    \end{multline}
for which we find the solution
    \begin{align}
        \mathcal B_{\mathbf C_{\widetilde{\mathbf{UV}}}}&=\Big(-q q_{V^T} z + 2 q z - q_{V^T}  \bar\sigma_X^2 - q_{V^T} z + z \nn\\ 
        &- \sqrt{q^2 q_{V^T}^2 z^2 - 2 q q_{V^T}^2  \bar\sigma_X^2 z + 2 q q_{V^T}^2 z^2 - 2 q q_{V^T} z^2 + q_{V^T}^2  \bar\sigma_X^4 + 2 q_{V^T}^2  \bar\sigma_X^2 z + q_{V^T}^2 z^2 - 2 q_{V^T}  \bar\sigma_X^2 z - 2 q_{V^T} z^2 + z^2}\Big)\nn\\
        &\Big/\big(2 z (q z - q_{V^T}  \bar\sigma_X^2 + z)\big)\,.
    \end{align}
Using Eq.~\eqref{eq:Rtransform_defn}, this gives us $\mathcal R_{\mathbf C_{\widetilde{{\mathbf U\mathbf V}}}}$, to which we add the $\mathcal R$-transform of the noise, Eq.~\eqref{eq:noise_Rtransform}, to obtain the $\mathcal R$-transform, $\mathcal R_\mathbf{C}$, of the NECM. Subtracting, $-1/z$, gives us the corresponding form of the $\mathcal B$-transform:
    \begin{align}
        \mathcal B_{\mathbf C}&=\Big(2 \widetilde\sigma^2 z (q z - q_{V^T}  \bar\sigma_X^2 + z) + (-q \widetilde\sigma^2 z + 1) (-q q_{V^T} z + 2 q z - q_{V^T}  \bar\sigma_X^2 - q_{V^T} z + z \nn\\&- \sqrt{q^2 q_{V^T}^2 z^2 - 2 q q_{V^T}^2  \bar\sigma_X^2 z + 2 q q_{V^T}^2 z^2 - 2 q q_{V^T} z^2 + q_{V^T}^2  \bar\sigma_X^4 + 2 q_{V^T}^2  \bar\sigma_X^2 z + q_{V^T}^2 z^2 - 2 q_{V^T}  \bar\sigma_X^2 z - 2 q_{V^T} z^2 + z^2})\Big)\nn\\
        &\Big/\big(2 z (-q \widetilde\sigma^2 z + 1) (q z - q_{V^T}  \bar\sigma_X^2 + z)\big)\,.
    \end{align}
The final step is to write down and solve the inverse functional equation $\mathcal B_{\mathbf C}[\mathfrak g_{\mathbf C}]=z$. The sixth order polynomial equation that we need to solve is of the form
\begin{align}\label{eq:intensive_polynomial}
        a\mathfrak g_{\mathbf C}^6+b\mathfrak g_{\mathbf C}^5+c\mathfrak g_{\mathbf C}^4+d\mathfrak g_{\mathbf C}^3+e\mathfrak g_{\mathbf C}^2+f\mathfrak g_{\mathbf C}+g=0\,,
\end{align}
with coefficients
\begin{align}
        a =& q^5 \widetilde\sigma^6 z^2 + 2 q^4 \widetilde\sigma^6 z^2 + q^3 \widetilde\sigma^6 z^2\,,\\
        b =& q^5 q_{V^T} \widetilde\sigma^6 z - 2 q^5 \widetilde\sigma^6 z - 2 q^4 q_{V^T} \bar\sigma_X^2 \widetilde\sigma^6 z^2 + 2 q^4 q_{V^T} \widetilde\sigma^6 z - q^4 \widetilde\sigma^6 z - 3 q^4 \widetilde\sigma^4 z^2 - 2 q^3 q_{V^T} \bar\sigma_X^2 \widetilde\sigma^6 z^2 + q^3 q_{V^T} \widetilde\sigma^6 z \nn\\&+ 3 q^3 \widetilde\sigma^6 z - 6 q^3 \widetilde\sigma^4 z^2 + 2 q^2 \widetilde\sigma^6 z - 3 q^2 \widetilde\sigma^4 z^2\,,\\
        c =& -q^5 q_{V^T} \widetilde\sigma^6 + q^5 \widetilde\sigma^6 - q^4 q_{V^T}^2 \bar\sigma_X^2 \widetilde\sigma^6 z + 3 q^4 q_{V^T} \bar\sigma_X^2 \widetilde\sigma^6 z - 3 q^4 q_{V^T} \widetilde\sigma^4 z - q^4 \widetilde\sigma^6 + 6 q^4 \widetilde\sigma^4 z + q^3 q_{V^T}^2 \bar\sigma_X^4 \widetilde\sigma^6 z^2 \nn\\&- q^3 q_{V^T}^2 \bar\sigma_X^2 \widetilde\sigma^6 z - 2 q^3 q_{V^T} \bar\sigma_X^2 \widetilde\sigma^6 z + 6 q^3 q_{V^T} \bar\sigma_X^2 \widetilde\sigma^4 z^2 + 2 q^3 q_{V^T} \widetilde\sigma^6 - 6 q^3 q_{V^T} \widetilde\sigma^4 z - 2 q^3 \widetilde\sigma^6 + 5 q^3 \widetilde\sigma^4 z + 3 q^3 \widetilde\sigma^2 z^2 \nn\\&- 4 q^2 q_{V^T} \bar\sigma_X^2 \widetilde\sigma^6 z + 6 q^2 q_{V^T} \bar\sigma_X^2 \widetilde\sigma^4 z^2 + q^2 q_{V^T} \widetilde\sigma^6 - 3 q^2 q_{V^T} \widetilde\sigma^4 z + q^2 \widetilde\sigma^6 - 5 q^2 \widetilde\sigma^4 z + 6 q^2 \widetilde\sigma^2 z^2 + q \widetilde\sigma^6 - 4 q \widetilde\sigma^4 z + 3 q \widetilde\sigma^2 z^2\,,\\
        d =& q^4 q_{V^T}^2 \bar\sigma_X^2 \widetilde\sigma^6 - q^4 q_{V^T} \bar\sigma_X^2 \widetilde\sigma^6 + 3 q^4 q_{V^T} \widetilde\sigma^4 - 3 q^4 \widetilde\sigma^4 - q^3 q_{V^T}^2 \bar\sigma_X^4 \widetilde\sigma^6 z - q^3 q_{V^T}^2 \bar\sigma_X^2 \widetilde\sigma^6 \nn\\&+ 3 q^3 q_{V^T}^2 \bar\sigma_X^2 \widetilde\sigma^4 z + 3 q^3 q_{V^T} \bar\sigma_X^2 \widetilde\sigma^6 - 9 q^3 q_{V^T} \bar\sigma_X^2 \widetilde\sigma^4 z + q^3 q_{V^T} \widetilde\sigma^4 + 3 q^3 q_{V^T} \widetilde\sigma^2 z + q^3 \widetilde\sigma^4 - 6 q^3 \widetilde\sigma^2 z + 2 q^2 q_{V^T}^2 \bar\sigma_X^4 \widetilde\sigma^6 z \nn\\&- 3 q^2 q_{V^T}^2 \bar\sigma_X^4 \widetilde\sigma^4 z^2 - q^2 q_{V^T}^2 \bar\sigma_X^2 \widetilde\sigma^6 + 3 q^2 q_{V^T}^2 \bar\sigma_X^2 \widetilde\sigma^4 z + 2 q^2 q_{V^T} \bar\sigma_X^2 \widetilde\sigma^4 z - 6 q^2 q_{V^T} \bar\sigma_X^2 \widetilde\sigma^2 z^2 - 4 q^2 q_{V^T} \widetilde\sigma^4 \nn\\&+ 6 q^2 q_{V^T} \widetilde\sigma^2 z + 5 q^2 \widetilde\sigma^4 - 7 q^2 \widetilde\sigma^2 z - q^2 z^2 - 2 q q_{V^T} \bar\sigma_X^2 \widetilde\sigma^6 + 8 q q_{V^T} \bar\sigma_X^2 \widetilde\sigma^4 z - 6 q q_{V^T} \bar\sigma_X^2 \widetilde\sigma^2 z^2 - 2 q q_{V^T} \widetilde\sigma^4 \nn\\&+ 3 q q_{V^T} \widetilde\sigma^2 z + q \widetilde\sigma^2 z - 2 q z^2 - \widetilde\sigma^4 + 2 \widetilde\sigma^2 z - z^2\,,\\
        e =& -3 q^3 q_{V^T}^2 \bar\sigma_X^2 \widetilde\sigma^4 + 3 q^3 q_{V^T} \bar\sigma_X^2 \widetilde\sigma^4 - 3 q^3 q_{V^T} \widetilde\sigma^2 + 3 q^3 \widetilde\sigma^2 - q^2 q_{V^T}^2 \bar\sigma_X^4 \widetilde\sigma^6 + 3 q^2 q_{V^T}^2 \bar\sigma_X^4 \widetilde\sigma^4 z + 2 q^2 q_{V^T}^2 \bar\sigma_X^2 \widetilde\sigma^4 \nn\\&- 3 q^2 q_{V^T}^2 \bar\sigma_X^2 \widetilde\sigma^2 z - 6 q^2 q_{V^T} \bar\sigma_X^2 \widetilde\sigma^4 + 9 q^2 q_{V^T} \bar\sigma_X^2 \widetilde\sigma^2 z - 2 q^2 q_{V^T} \widetilde\sigma^2 - q^2 q_{V^T} z + q^2 \widetilde\sigma^2 + 2 q^2 z + q q_{V^T}^2 \bar\sigma_X^4 \widetilde\sigma^6 \nn\\&- 4 q q_{V^T}^2 \bar\sigma_X^4 \widetilde\sigma^4 z + 3 q q_{V^T}^2 \bar\sigma_X^4 \widetilde\sigma^2 z^2 + 2 q q_{V^T}^2 \bar\sigma_X^2 \widetilde\sigma^4 - 3 q q_{V^T}^2 \bar\sigma_X^2 \widetilde\sigma^2 z - 2 q q_{V^T} \bar\sigma_X^2 \widetilde\sigma^4 + 2 q q_{V^T} \bar\sigma_X^2 \widetilde\sigma^2 z \nn\\&+ 2 q q_{V^T} \bar\sigma_X^2 z^2 + 2 q q_{V^T} \widetilde\sigma^2 - 2 q q_{V^T} z - 3 q \widetilde\sigma^2 + 3 q z + 2 q_{V^T} \bar\sigma_X^2 \widetilde\sigma^4 - 4 q_{V^T} \bar\sigma_X^2 \widetilde\sigma^2 z + 2 q_{V^T} \bar\sigma_X^2 z^2 + q_{V^T} \widetilde\sigma^2 \nn\\&- q_{V^T} z - \widetilde\sigma^2 + z\,,\\
        f =& 3 q^2 q_{V^T}^2 \bar\sigma_X^2 \widetilde\sigma^2 - 3 q^2 q_{V^T} \bar\sigma_X^2 \widetilde\sigma^2 + q^2 q_{V^T} - q^2 + 2 q q_{V^T}^2 \bar\sigma_X^4 \widetilde\sigma^4 - 3 q q_{V^T}^2 \bar\sigma_X^4 \widetilde\sigma^2 z - q q_{V^T}^2 \bar\sigma_X^2 \widetilde\sigma^2 + q q_{V^T}^2 \bar\sigma_X^2 z \nn\\&+ 3 q q_{V^T} \bar\sigma_X^2 \widetilde\sigma^2 - 3 q q_{V^T} \bar\sigma_X^2 z + q q_{V^T} - q - q_{V^T}^2 \bar\sigma_X^4 \widetilde\sigma^4 + 2 q_{V^T}^2 \bar\sigma_X^4 \widetilde\sigma^2 z - q_{V^T}^2 \bar\sigma_X^4 z^2 - q_{V^T}^2 \bar\sigma_X^2 \widetilde\sigma^2 + q_{V^T}^2 \bar\sigma_X^2 z \nn\\&+ 2 q_{V^T} \bar\sigma_X^2 \widetilde\sigma^2 - 2 q_{V^T} \bar\sigma_X^2 z\,,\\
        g =& -q q_{V^T}^2 \bar\sigma_X^2 + q q_{V^T} \bar\sigma_X^2 - q_{V^T}^2 \bar\sigma_X^4 \widetilde\sigma^2 + q_{V^T}^2 \bar\sigma_X^4 z\,.
    \end{align}
\end{widetext}
We solve this polynomial equation numerically, looking for complex roots which yield non-zero values of the eigenvalue density. For large signal-to-noise ratio, we encounter numerical instabilities trying to determine the eigenvalue density bounds. We run into these instabilities in the determination of the true density bounds in Fig.~\ref{fig:fig2}(d). To fix this, we start at the peak of the density and determine the values of $\lambda$, for which it hits zero for the first time, to either side of the peak. All other zero density crossings are assumed to be due to numerical instabilities. 

The ranges of nonzero density are shown in Fig.~\ref{fig:fig2}(d). We see that, as the signal-to-noise ratio increases, there is a bifurcation point where the density splits into two bumps. The left bump is associated with the noise, and the right bump is associated with pure latent feature signal. From our approximate expression for the eigenvalue bounds,  Eq.~\eqref{eq:bound_approx}, we can estimate the value of the $\textsf{SNR}$ at which the splitting occurs. The defining equation for this  is given by the intersection between the right boundary of the noise region and the left boundary of the signal part of the densit in Eq.~\eqref{eq:bound_approx}, resulting in:
\begin{align}
    (1+\textsf{SNR}^{-1})\lambda_+^\text{MP}=\lambda_-^\textsf{SNR}\,.
\end{align}
Solving for $\textsf{SNR}$, we obtain the following estimation for the splitting point:
\begin{align}\label{eq:SNR_bifurcation}
    \textsf{SNR}_\text{split}\approx\frac{\lambda_+^\text{MP}-\lambda_-^\text{MP}}{\lambda_-^{\infty}}\,.
\end{align}


\providecommand{\noopsort}[1]{}\providecommand{\singleletter}[1]{#1}%

\end{document}